\documentclass[twocolumn]{aastex62}

\graphicspath{{./}{figures/}}
\usepackage[utf8]{inputenc}
\usepackage{graphicx}% Include figure file
\usepackage{dcolumn}% Align table columns on decimal point
\usepackage{bm}% bold math
\usepackage{ulem} %here for \sout, removed to avoid funny underlines
\usepackage{booktabs} 
\usepackage{hyperref}% add hypertext capabilities
\usepackage{bm}
\usepackage{multirow,enumitem}
\usepackage{amsmath}
\usepackage{color}
\usepackage{xcolor}
\usepackage{multirow}

\newcommand{\LCDM}{\rm{\Lambda}CDM}

\newcommand{\Mpc}{\mathrm{~km~s^{-1}~Mpc^{-1}}}

\begin{document}

\title{Breaking the Mass-sheet Degeneracy in Time-delay Cosmography with Lensed and Unlensed Type Ia Supernovae}

\author{Xiaolei Li}
\affiliation{College of Physics, Hebei Normal University, Shijiazhuang 050024, China}
\affiliation{Shijiazhuang Key Laboratory of Astronomy and Space Science, Hebei Normal University, Shijiazhuang, Hebei 050024, China}
\author{Kai Liao}\thanks{liaokai@whu.edu.cn}
\author{Xuheng Ding}
\affiliation{School of Physics and Technology, Wuhan University, Wuhan 430072, China}

\date{\today}% It is always \today, today,
 
\begin{abstract}

This study introduces an innovative framework aimed at overcoming the ongoing issue of mass-sheet degeneracy (MSD) in time-delay cosmography by incorporating observations of both gravitationally lensed and unlensed Type Ia supernovae (SNe Ia). By simultaneously using lensing magnification measurements $\mu^{\rm{obs}}$ and cosmic distance ratios
($D_s/D_{ds}$), we develop a Bayesian framework capable of breaking the MSD. Specifically, we reconstruct the distance-redshift and magnitude-redshift relations from unlensed Type Ia supernovae using Gaussian Process to avoid dependence on specific cosmological models.
Our framework shows substantial efficacy in resolving the MSD by imposing constraints on the MSD parameter $\lambda$. Furthermore, we extend this framework to analyze multiple gravitational lensing systems. The results show strong agreement with the fiducial MSD parameters used in the data simulation, confirming our method's effectiveness in mitigating the MSD. Ultimately, this technique enables the derivation of corrected time-delay distance measurements under the MSD, improving the precision of cosmological parameters inferred from strong lensing systems.

\end{abstract}

%Use showkeys class option if keyword
                              %display desired
% \maketitle

\section{Introduction}

The precise determination of cosmological parameters, particularly the Hubble constant $H_0$ that quantifies the current cosmic expansion rate, represents a critical frontier in modern cosmology \citep{Planck:2018vyg, Riess_2022}. Although early-universe constraints from cosmic microwave background (CMB) anisotropies \citep{Planck:2018vyg} and late-universe calibrations via Type Ia supernovae \citep{Riess_2022} have percent-level precision, their persistent discrepancy (the so-called "Hubble tension") now exceeds $5\sigma$ significance \citep{DiValentino:2021izs}.This issue highlights the imperative necessity for independent geometric methodologies of measurement $H_0$, with Time-Delay cosmography emerging as an especially promising technique \citep{Refsdal:1964blz, Treu:2016ljm,Millon:2019slk,TDCOSMO:2025dmr}.

The  Time-Delay cosmography method leverages Fermat's principle in general relativity: light rays from a background source traverse different paths around a massive lens, creating multiple images with time delays proportional to the Hubble constant \citep{Refsdal1964MNRAS,Treu:2010uj,Liao:2022gde}. Modern analyses achieve $\sim 2\%$ precision per lens system through exquisite modeling of lens mass distributions \citep{H0LiCOW:2019pvv, Millon:2019slk,Li:2024elb}.
Nonetheless, this precision is inherently limited by the mass-sheet degeneracy (MSD), which is a mathematical transformation that maintains lensing observables while modifying cosmological inferences \citep{1985ApJ...289L...1F,Schneider:2013sxa,Treu:2022aqp}.

MSD is a fundamental limitation in  Time-Delay cosmography. It describes an inherent ambiguity in lens modeling where multiple distinct mass distributions can reproduce the same observational data, e.g., the image positions and magnification ratios of lensed sources. Mathematically, given an original lens model with convergence $\kappa(\theta)$ and source position $\beta$, there exists a family of transformed models parameterized by $\lambda$
\begin{equation}
\kappa_{\lambda} = (1 - \lambda) + \lambda \kappa; \quad \beta_{\lambda} = \beta / \lambda,
\end{equation}
which preserve the observed image configurations but alter the predicted time delays
\begin{equation}
\Delta t_{\lambda} = \lambda \Delta t,
\end{equation}
and magnifications 
\begin{equation}
\quad \mu_{\lambda} = \mu / \lambda^2.
\end{equation}
This degeneracy arises because a uniform "mass sheet" (scaled by $\lambda$)
can be added to the lens model without changing the relative image properties, while systematically biasing absolute quantities like the Hubble constant inferred from time-delay cosmography.
The MSD poses a significant challenge for precision cosmology, as it requires external constraints (e.g., independent distance measurements or lens kinematics) to break the degeneracy and derive unbiased cosmological parameters. 
% As an intrinsic transformation preserving the invariance of lensing observables while modifying time delay measurements, MSD constitutes a fundamental and persistent challenge in the analysis of strong gravitational lensing. This issue is especially significant when attempting to constrain the precision of $H_0$ measurements in the absence of additional data.

% Formally, this transformation modifies the lensing convergence $\kappa$ as:
% \begin{equation}
%   \kappa_{\lambda}(\boldsymbol{\theta}) = \lambda \kappa(\boldsymbol{\theta}) + (1 - \lambda),
% \end{equation}
% where $\lambda$ is a constant scaling factor, and $\boldsymbol{\theta}$ denotes angular coordinates on the sky. This rescaling preserves the relative gravitational potential differences that govern time delays but alters the absolute mass normalization of the lens. Consequently, the inferred cosmological distances (and thus parameters like $H_0$) become degenerate with the unknown MSD parameter $\lambda$. For example, a positive external convergence (e.g., from large-scale structure or nearby galaxy groups) mimics the effect of a mass sheet, biasing the inferred $H_0$ low if unaccounted for, while a negative convergence (underdense regions) biases $H_0$ high. In order to get a reliable estimate of $H_0$ from  Time-Delay cosmography, one first needs to break the MSD. However, this degeneracy is particularly pernicious because it cannot be resolved using lensing data alone, even with exquisite measurements of image configurations and time delays.

Various methodologies have been explored to overcome the MSD. For example, the TDCOSMO collaboration, as noted in \cite{Birrer:2020tax}, introduced a hierarchical framework to limit the effects of the MSD using stellar kinematics data from the deflector galaxy population, effectively minimizing covariances among individual lens assumptions. By combining 7 TDCOSMO strong lenses with 33 galaxy-galaxy lenses from the SLACS survey, a precision measurement of $5\%$ was achieved at $H_0$. Projections for future constraints on breaking the MSD using kinematic observations under analogous analytical assumptions have been provided in \cite{Birrer:2020jyr}. 
{Building on the progress, the latest analysis by \cite{TDCOSMO:2025dmr} further advanced the breaking of the MSD by expanding the sample of time-delay lenses to eight and, more importantly, incorporating significantly improved spatially resolved stellar kinematics measurements. These enhanced data provided stronger constraints on the mass distribution, enabling a more robust breaking of the MSD and yielding a refined $H_0$ measurement with a precision of $4–5\%$.}
% Nevertheless, kinematic measurements remain subject to the mass-anisotropy degeneracy, whose mitigation requires either the adoption of assumptions regarding the stellar anisotropy distribution or the acquisition of spatially resolved kinematic data. \cite{Birrer:2020jyr} subsequently provided projections data for amelioration. 

Furthermore, the study by \cite{Chen:2020knz} employed the distance ratio $D_s/D_{ds}$ to provide the stellar velocity dispersion of the lensing galaxy. The distance ratio can be derived either by assuming a particular cosmological model or by utilizing relative distance indicators such as SNe Ia and Baryon Acoustic Oscillations (BAO) to acquire a cosmological-model-independent distance ratio. In their research, the authors utilized Piece-wise Natural Cubic Splines to describe $H(z)$, which were fitted to the data to derive the redshift-distance relationship from external distance indicators. This spline approach is responsive to data at high redshifts and may introduce bias into the spline fit, thereby transmitting errors to the distance ratio. Additionally, if the covariance between different distance indicators is disregarded, the uncertainties may be underestimated.

In the study conducted by \cite{Birrer:2021use}, a hierarchical Bayesian analysis framework was introduced to resolve the MSD by combining a dataset of unlensed SNe Ia with measured apparent magnitudes and another dataset of strongly lensed SNe Ia, which includes measured relative time delays and apparent magnitudes. To determine the absolute luminosity of an unlensed SN Ia at the redshift corresponding to the lensed SNe Ia, the study employed the apparent magnitude of the unlensed SNe Ia obtained from the cosmological model sample. This was inferred at the redshift of the lensed SN Ia by replacing the absolute magnitude $M_B$ term with an apparent magnitude at a specific redshift, as outlined in $z_{\rm{pivot}}$ and $m_p$. However, this method of inferring the absolute luminosity of unlensed SNe Ia at lensed SNe Ia redshifts is contingent upon the redshift-independent peak brightness assumption. This assumption may not be valid for high-redshift SNe Ia due to potential evolutionary effects. Additionally, the correlations of magnitudes across different redshifts of SNe Ia are not taken into account.

{To break MST, one need to know the true $D_s/D_{ds}$  prior either by assuming a cosmological model \citep{2020A&A...643A.165B} or from unlensed SNe Ia \cite{Chen:2020knz}, with the following relation:
\begin{equation}
(\sigma_\nu^P)^2=(\frac{D_s}{D_{ds}})_{true}J(\eta_{lens},\eta_{light},\beta_{ani},\lambda),
\end{equation}
with
\begin{equation}
J(\lambda)\approx\lambda J.
\end{equation}}

As lensed supernovae receive increasing attention \citep{2024SSRv..220...13S, Liao:2022gde,2019RPPh...82l6901O}, this study presents a novel framework to break the MSD by combining unlensed and gravitationally lensed SNe Ia observations. Our approach employs {Gaussian Process (GP) }regression to reconstruct both the distance-redshift and magnitude-redshift relations in a cosmology-independent manner, enabling precise determination of distance ratios for lens systems and source magnitudes at their respective redshifts. Crucially, this method fully utilizes the entire SNe Ia sample to minimize uncertainties at respective redshifts through optimal information extraction.
We will consider two methodologies to break the MSD simultaneously. We are going to use magnification constraints from flux comparisons between lensed and unlensed SNe Ia, as well as distance ratio measurements derived from the reconstructed relations, to break the MSD. This combination significantly enhances the robustness of breaking MSD compared to single-method analyses. Furthermore, we develop a hierarchical Bayesian system capable of analyzing multiple time-delay lens systems collectively, substantially improving the statistical power to constrain the MSD parameter $\lambda$.
{We clarify that our goal is not to introduce a new use for stellar kinematics, but to correct the bias in lens models that ignore internal MSD. Quantities like $D_{\Delta T}^{PL}$ and $(D_s/D_{ds})^{PL}$ are biased. Breaking the MSD requires an independent constraint on the true distance ratio, $(D_s/D_{ds})_{\rm true}$}.   
By incorporating the posterior distribution of the MSD parameter along with the time-delay distance, unbiased measurements of the time-delay distance can be achieved. This approach effectively removes systematic biases caused by the MSD, thereby enabling more accurate and reliable estimation of cosmological parameters. Our methodology is particularly valuable for next-generation surveys (e.g., Rubin and Roman observatories) that will discover large samples of lensed SNe Ia.

This work is organized as follows: In Sec.~\ref{sec:data}, we briefly describe the numerical simulations and data generation process, including the simulations of unlensed SNe Ia data, data reconstruction with GP as well as the simulations and estimations of glSNe assuming a power-law lens model. In Sec.~\ref{sec:method}, we show the novel Bayesian formalism which is developed to break MSD in individual glSNe systems through joint likelihood analysis of magnifications and time-delay distances, with Markov Chain Monte Carlo (MCMC) implementations for posterior sampling. 
In Sec.~\ref{sec:multi-systems}, we demonstrate our correlation-aware methodology through analysis of two lensing systems, illustrating its application in multi-system scenarios. And in Sec.~\ref{sec:time-delayDistance}, we show the unbiased time-delay distance under the MSD. The final conclusions are presented in Sec.~\ref{sec:con}.

%Recent advances in combining  Time-Delay cosmography with independent distance indicators, such as standard candles (e.g., Type Ia supernovae, SNe Ia), offer a promising pathway to break this degeneracy. In the work of \cite{Birrer:2021use}, the authors break the MSD with the observations of Strongly lensed supernovae (glSNe) based on the fact that glSNe can provide, in addition to measurable time delays, lensing magnification constraints when knowledge about the unlensed apparent brightness of the explosion is imposed .

% This work explores the theoretical framework and observational feasibility of integrating standard candles into  Time-Delay cosmography analyses. We demonstrate how joint modeling of time-delay lenses and SNe Ia datasets can yield robust constraints on both the lens mass profile and $H_0$, while quantifying the reduction in systematic uncertainties from the MSD.

\section{Data}\label{sec:data}
Since there is a lack of necessary lensed SNe Ia observations, we will simulate the lensed SNe Ia systems based on a fiducial cosmological model. Moreover, to make the results convincing, we use the simulated unlensed SNe Ia instead of the real data from observations.

In this section, we briefly describe the numerical simulations and data generation process. All cosmological simulations in this work adopt a spatially flat $\LCDM$ framework as the fiducial model, with the matter density parameter $\Omega_M = 0.3$ and Hubble constant $H_0 = 70\ \mathrm{km\,s^{-1}\,Mpc^{-1}}$. The cosmological parameters remain fixed throughout our analysis to ensure consistency in degeneracy-breaking evaluations.

\subsection{Configuration for lens model}
\begin{figure*}
\centering
\includegraphics[width=0.75\textwidth]{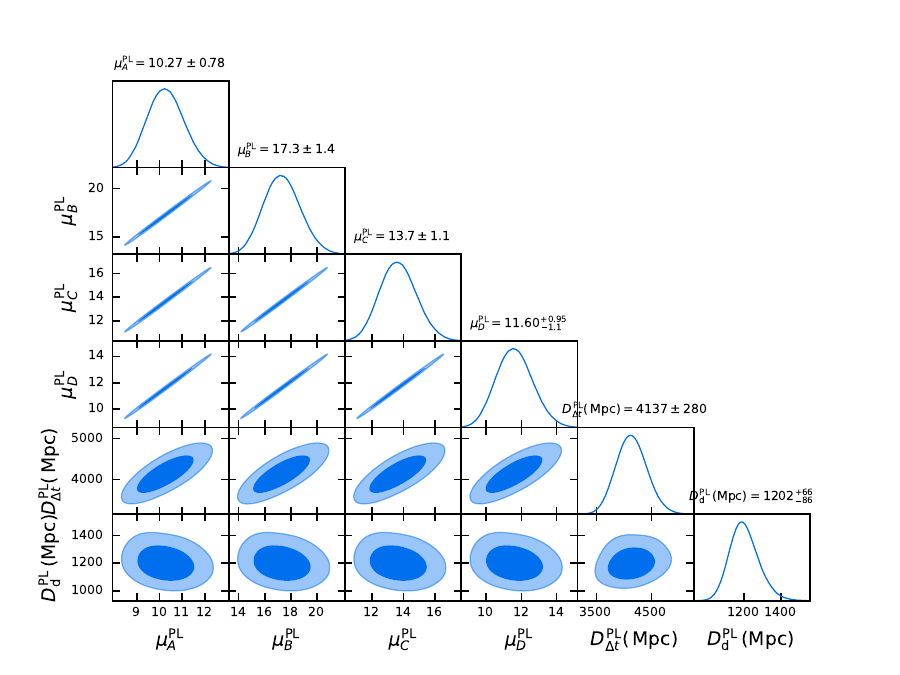}
\caption{The time-delay distance ($D_{\Delta t}$) and angular-diameter distance to the lens ($D_{\rm{d}}$), along with the magnification ratios of the lensed images. The results are derived from a power-law lens model in \textsc{Lenstronomy}, with the incorporation of an intrinsic MSD parameter.}
  \label{Fig:muiddt}%
\end{figure*}

Assuming a power-law lens model featuring the quad lensing system, we utilize \textsc{Lenstronomy} \citep{Birrer:2021use} to estimate two crucial observables for a gravitational lensing system, situated in the context of the fiducial cosmological model.
% : the distances, notably the time-delay distance $D_{\Delta t}^{\rm fid}$ and the angular diameter distance to the lens galaxy $D_d^{\rm fid}$, alongside the magnifications of the four images, $\mu_i$ where $i = A, B, C, D$ represents the four images. 
The configurations and observational parameters pertinent to this lensing system are displayed in the third column of Table~\ref{tab:lenstronomy}, referred to as Lens 01.

{$D_{\Delta T}^{\mathrm{PL}}$ and $D_d^{\mathrm{PL}}$ as well as $\mu_i^{\mathrm{PL}}$ are got following the standard pipeline \textsc{Lenstronomy} \citep{Birrer:2021use}. The equations are:
\begin{equation}
(\sigma_\nu^P)^2=(1-\kappa_{\mathrm{ext}})\left(\frac{D_s}{D_{ds}}\right)^{\mathrm{PL}}J(\eta_{\mathrm{lens}},\eta_{\mathrm{light}},\beta_{\mathrm{ani}}),
\end{equation}
and 
\begin{equation}
D_d=\frac{1}{1+z_d}D_{\Delta T}^{\mathrm{PL}}\frac{c^2}{(\sigma_\nu^P)^2}J(\eta_{\mathrm{lens}},\eta_{\mathrm{light}},\beta_{\mathrm{ani}})
\end{equation}
We didn't consider the MSD parameter $\lambda$ in $J(\eta_{\mathrm{lens}},\eta_{\mathrm{light}},\beta_{\mathrm{ani}})$, thus the $D_{\Delta T}^{\mathrm{PL}}$ and $(D_s/D_{ds})^{PL}$ as well as $\mu_i^{\mathrm{PL}}$ are actually biased due to the MSD (Note the $\lambda$ is approximately canceled out in $D_d$ which is not biased.)}

% \textbf{$D_{\Delta T}^{\rm PL}$ and ${D_d^{{\rm PL}}}$ as well as $\mu_i^{\rm PL}$ are got following the standard pipline in H0LiCOW or Time Delay Lens Modelling Challenge program （TDLMC). We didn't consider the MSD parameter $\lambda$. The equations are:
% \begin{equation}
% (\sigma_\nu^P)^2=(1-\kappa_{ext})(\frac{D_s}{D_{ds}})^{\rm PL}J(\eta_{lens},\eta_{light},\beta_{ani}),
% \end{equation}
% and 
% \begin{equation}
% D_d=\frac{1}{1+z_d}D_{\Delta T}^{\rm PL}\frac{c^2}{(\sigma_\nu^P)^2}J(\eta_{lens},\eta_{light},\beta_{ani})
% \end{equation}
% Thus the $D_{\Delta T}^{\rm PL}$ and ${D_d^{{\rm PL}}}$ as well as $\mu_i^{\rm PL}$ are actually biased due to the MSD.
% }
% Moreover, we add an intrinsic MSD parameter $\lambda^{\rm fid} = 1.2$ to the results of fiducial cosmological model following
% \begin{equation}
%     D_{\Delta t}^{\rm PL} = D_{\Delta t}^{\rm fid}\lambda^{\rm fid}, D_{d}^{\rm PL} = D_{d}^{\rm fid},
% \end{equation}
% and 
% \begin{equation}
%     \mu_{i}^{\rm PL} = \mu_{i}^{\rm fid}(\lambda^{\rm fid})^2.
% \end{equation}
According to the measurement precision level for lensed quasars and considering the improvements for lensed supernovae as transient sources \citep{2017NatCo...8.1148L,2021MNRAS.504.5621D}, we incorporate realistic lensing model constraints on key parameters, accounting for the error budget in the mass slope, point source astrometry, time delay, velocity dispersion, and external convergence (see Table~\ref{tab:lenstronomy}); and we employ random simulations to generate corresponding estimates of the inferred parameters, including the magnifications of the four lensed images, $D_{\Delta t}^{\rm PL}$ and $D_{d}^{\rm PL}$. The final estimation of these parameters is shown in Figure~\ref{Fig:muiddt}.
{It is true that $D_{\Delta T}^{\rm PL}$ and ${D_d^{\rm PL}}$ should be correlated and we had considered this correlation in this work. However, the correlation is not be as strong as expected for each system. Similar weak correlation has been reported in \cite{TDCOSMO:2023hni,H0LiCOW:2019pvv,H0LiCOW:2018tyj}.
In \cite{Jee:2015yra}, the authors even ignored the correlation for simplicity. }

{For full methodological and implementational details, we refer the readers to the foundational \textsc{Lenstronomy} publications, including \cite{Birrer:2018xgm, Ding:2020jmg}, and the official code repository: \href{https://github.com/lenstronomy}{\texttt{github.com/lenstronomy}}.}

We apply this lensing configuration and its corresponding estimations to our simulated lensed SNe Ia systems, as described later in Section~\ref{sec:lens_SNe}.

\begin{table*}[h]
\centering
\caption{Simulation configurations and observational parameters for the two lensing systems.}\label{tab:lenstronomy}
\begin{tabular}{llcc}
\hline
Parameter & Symbol & Lens 01 & Lens 02 \\
\midrule
\multicolumn{4}{l}{\textbf{Cosmology and redshifts}} \\
Lens redshift & $z_{\text{lens}}$ & 0.5 & 0.5 \\
Source redshift & $z_{\text{source}}$ & 1.5 & 1.0 \\
Hubble constant & $H_0\, (\Mpc)$ & {70} & {70} \\
Matter density & $\Omega_m$ & 0.3 & 0.3 \\
\hline
\multicolumn{4}{l}{\textbf{Lens model parameters}} \\
Power-law slope & $\gamma$ & 2.0 & 2.2 \\
Einstein radius & $\theta_E\,(^{\prime \prime})$ & 1.66 & 1.9 \\
Shear $\gamma_1$ & $\gamma_1$ & 0.06 & 0.06 \\
Shear $\gamma_2$ & $\gamma_2$ & -0.02 & -0.02 \\
Ellipticity $e_1$ & $e_1$ & 0.05 & 0.05 \\
Ellipticity $e_2$ & $e_2$ & 0.05 & 0.05 \\
Source position $x$ & $\beta_x\,(^{\prime \prime})$ & -0.05 & -0.25 \\
Source position $y$ & $\beta_y\,(^{\prime \prime})$  & 0.02 & 0.17 \\
MSD parameter & $\lambda$  & 1.2 & 0.9\\
\hline
\multicolumn{4}{l}{\textbf{Observational parameters}} \\
Slit length & $R_{\text{slit}}\,(^{\prime \prime})$ & 1.0 & 1.0 \\
Slit width & $dR_{\text{slit}}\,(^{\prime \prime})$  & 1.0 & 1.0 \\
PSF FWHM & FWHM $(^{\prime \prime})$  & 0.7 & 0.7 \\
Anisotropy model & - & OM & OM \\
Anisotropy radius & $r_{\text{ani}}$ & 1.0 & 1.0 \\
\hline
\multicolumn{4}{l}{\textbf{Measurement errors}} \\
Mass slope error & $\sigma_\gamma$ & 0.03 & 0.03 \\
{Astrometry error} & $\sigma_{\text{astrometry}} \,(^{\prime \prime})$& 0.003 & 0.003 \\
{Time delay error} & $\sigma_{\Delta t}$ (day) & 1.0 & 1.0 \\
{Vel. dispersion error} & $\sigma_{v_{\text{disp}}}$ (${\mathrm{~km~s^{-1}}}$) & 10.0 & 10.0 \\
Ext. convergence error & $\sigma_{\kappa_{\text{ext}}}$ & 0.02 & 0.02 \\
\hline
\end{tabular}
\end{table*}

% The corresponding magnification factors for the multiple lensed images are presented in Figure~\ref{Fig:mag}.

% \begin{figure}
% \centering
% \includegraphics[width=0.5\textwidth]{Lens01-mu4PL.pdf}
% \caption{The estimated magnifications for the four images based on a power-law lens model with $z_{\rm{s}} = 1.5$, $z_{\rm{d}}=0.5$.}
%   \label{Fig:mag}%
% \end{figure}

\subsection{Unlensed SNe Ia simulations}

Within the framework of the fiducial cosmological model, we generate synthetic SNe Ia observations using the Pantheon+ sample as our baseline dataset. The Pantheon+ sample is a comprehensive dataset of 1,701 spectroscopically confirmed SNe Ia, which is a significant expansion over its predecessor, the Pantheon sample (1,048 SNe Ia). This dataset is one of the most precise collections of SNe Ia used for cosmological studies, particularly in measuring the expansion history of the universe and constraining dark energy parameters \citep{Brout:2022vxf}.
First, we obtain the luminosity distances of SNe Ia with
\begin{equation}\label{eq:DL}
    D_{\rm{L}}^{\rm{fid}}(z)\,=\,\frac{c(1+z)}{H_0} \int_0^z \frac{dz}{\sqrt{\Omega_m(1+z)^3+(1-\Omega_m)}}
\end{equation}
and then the magnitude can be calculated with
\begin{equation}\label{eq:mfidunlensed}
    m^{\rm{fid}}_{\rm{unlensed}}(z) \,=\, M+5\log_{10}{\frac{D^{\rm{fid}}_L(z)}{10}}
\end{equation}
where M = -19.3 represents the absolute magnitude of SN Ia \citep{Riess_2022}.
The simulated SNe Ia data, $m^{\rm{obs}}_{\text{unlensed}}(z)$, are then generated from $m^{\rm{fid}}_{\rm{unlensed}}(z)$ by adding noise as a random variable with a mean of zero and a variance characterized by the Pantheon+ covariance matrix. 

\begin{figure}
\centering
\includegraphics[width=0.5\textwidth]{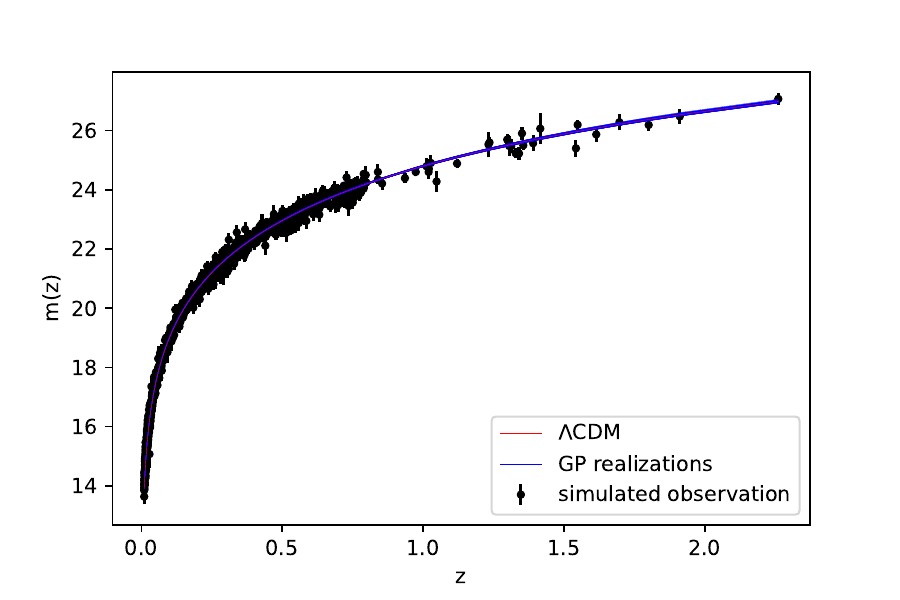}
\caption{The simulated observation of magnitudes $m^{\rm{obs}}_{\rm{unlensed}}$ and the reconstructed magnitudes $m^{\rm{GP}}_{\rm{unlensed}}$ .}
  \label{Fig:mzunlensed}%
\end{figure}

\begin{figure}
\centering
\includegraphics[width=0.5\textwidth]{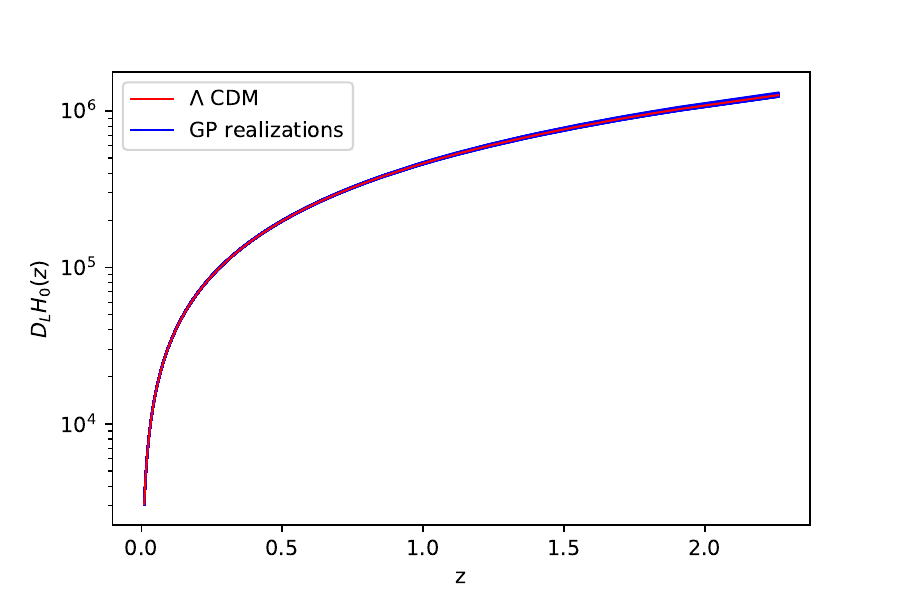}
\caption{The reconstructed unanchored luminosity distance $(D_LH_0)^{\rm{GP}}_{\rm{unlensed}}$ from simulated distance module based on flat $\LCDM$ model.}
  \label{Fig:DLH0unlensed}%
\end{figure}

The unlensed magnitude measurements $m^{\rm{obs}}_{\rm{unlensed}}$ and the unanchored luminosity distance $(D_LH_0)^{\rm{GP}}_{\rm{unlensed}}$ are reconstructed from the simulated SNe Ia data set using GP. 
GP works by generating a random set of cosmological functions whose statics are characterized by a covariance function. We follow some previous works and use a squared-exponential kernel for the covariance function ~\citep{Li:2021onq,Li:2023gpp,Li:2024elb,Li:2024hed} 
\begin{equation}\label{eq:kernel}
   <\varphi(s_i)\varphi(s_j)>\,=\,\sigma_f^2 \exp\left({-\frac{|s_i-s_j|^2}{2\ell^2}}\right)
\end{equation}
where $s_i=\ln (1+z_i)/\ln (1+z_{\rm{max}})$ and $z_{\rm{max}}=2.261$ is the maximum redshift of the SNe Ia sample.  $\sigma_f$ and $\ell$ are two hyperparameters that are marginalized.
$\varphi$ is just a random function drawn from the distribution defined by the covariance function of Equation~(\ref{eq:kernel}) and we take this function as $\varphi(z)=\ln \left(H^{\rm{mf}}(z)/H(z)  \right)$, i.e. the logarithm of the ratio between the reconstructed expansion history, $H(z)$, and a mean function, $H^{\rm{mf}}(z)$, which we choose to be the best-fit $\LCDM$ model from the Pantheon+ data set. All quantities with a `GP' subscript or superscript are derived from this GP reconstruction and therefore carry statistical uncertainties; they are treated as observables in the likelihood. The results are shown in Figure~\ref{Fig:mzunlensed} for magnitude reconstructions and Figure~\ref{Fig:DLH0unlensed} for unanchored luminosity distance, respectively. {Since both the reconstructed unlensed magnitude $m^{\rm{obs}}{\rm{unlensed}}$ and the unanchored luminosity distance $(D_L/H_0)^{\rm{GP}}_{\rm{unlensed}}$ originate from the same SNe Ia data, they are statistically correlated. We explicitly incorporate this dependence via the hyperparameters in GP.}

{The GP regression is designed to be model-independent. Although we adopt a flat $\Lambda$CDM model as a mean function, the data-optimized hyperparameters are designed to capture all significant deviations from it, ensuring the final distance-redshift relation is data-driven \citep{Shafieloo2012Gaussian,2013PhRvD..87b3520S,2017JCAP...09..031A}. Consequently, $\Lambda$CDM acts merely as a computational prior and does not rigidly constrain the cosmological result.} For the details of the reconstruction with GP, we refer the readers to \citep{Rasmussen:2006,Holsclaw:2010nb,Holsclaw:2010sk,2011PhRvD..84h3501H,Shafieloo2012Gaussian,2013PhRvD..87b3520S,2017JCAP...09..031A,Keeley:2020aym,Li:2021onq,2023JCAP...02..014H}. 


\subsection{Lensed SNe Ia simulations}\label{sec:lens_SNe}
Based on the fiducial cosmological model, we calculate the unlensed apparent magnitude of a SN Ia at source redshift $z_s = 1.5$ with equation~(\ref{eq:DL}) and equation~(\ref{eq:mfidunlensed}).
% through:
% \begin{equation}
%     D_L^{\rm{fid}} (z_s)\,=\,\frac{c}{H_0}(1+z)\int_0^{z_s}\frac{d z}{E(z)}
% \end{equation}
% where the dimensionless Hubble parameter is defined as $E(z) = \sqrt{\Omega_M(1+z)^3+(1-\Omega_M)}$.
% The unlensed apparent magnitude of a SN Ia is then determined using
% \begin{equation}
%     m^{\rm{fid}}_{\rm{unlensed}}(z_s) \,=\, M+5\log_{10}{\frac{D^{\rm{fid}}_L(z_s)}{10}}
% \end{equation}
% where M = -19.3 represents the absolute magnitude of SN Ia.
The lensed apparent magnitude is subsequently derived from
\begin{equation}
  m^{\rm{fid}}_{\rm{lensed}}(z_s)\,=\,m^{\rm{fid}}_{\rm{unlensed}}(z_s)-2.5\log_{10}({\mu^{\rm{fid}}})
\end{equation}
where $\mu^{\rm{fid}} $ denotes the fiducial magnification factors of the lensed images obtained in \textsc{Lenstronomy}.

\begin{figure}
\centering
\includegraphics[width=0.5\textwidth]{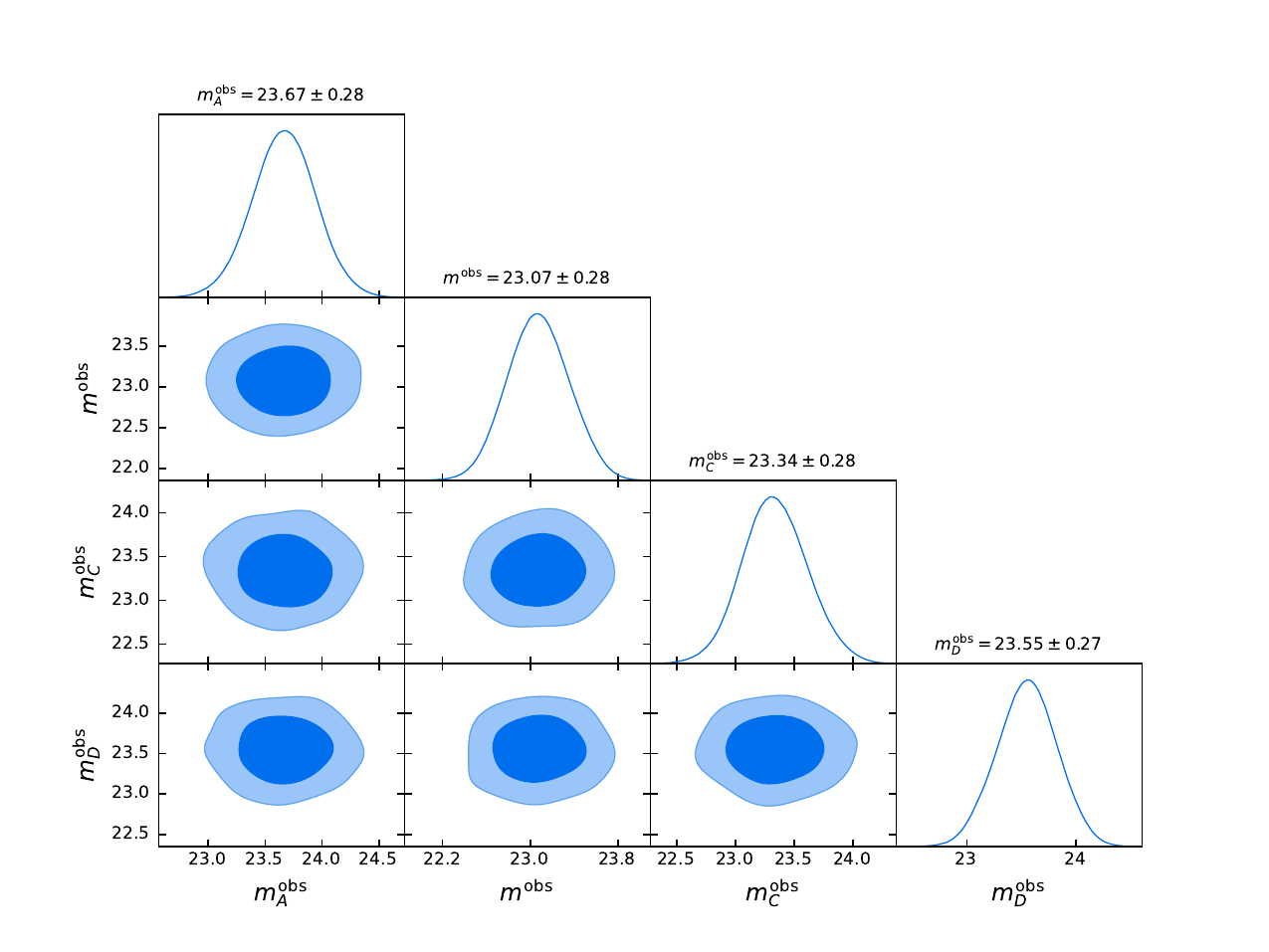}
\caption{Simulated lensed magnitudes $m^{\rm{obs}}_{\rm{lensed}}(z_s)$ with a total error of 0.28 mag.}
  \label{Fig:mzlensed}%
\end{figure}

We account for a total photometric error budget of 0.28 magnitudes in our simulations, incorporating three key error components: Intrinsic scatter (0.163 mag, corresponding to $15\%$ flux variation), Millilensing effects (0.11 mag, $10\%$ flux variation) and Microlensing uncertainties (constrained to $\leq 0.20$ mag, $20\%$ flux variation). These error terms are applied to the fiducial lensed magnitudes $m^{\mathrm{fid}}_{\mathrm{lensed}}(z_s)$ to generate realistic observational data, producing the simulated observed magnitudes $m^{\mathrm{obs}}_{\mathrm{lensed}}(z_s)$. The resulting distributions of these simulated lensed magnitude observations are shown in Figure~\ref{Fig:mzlensed}.

% We consider a total photometric error of 0.28 magnitudes, including Intrinsic scatter (about 0.163 mag), Millilensing (0.11 mag) and micorlensing (about 0.20 mag) error contributions to the intrinsic lensed magnitudes $m^{\mathrm{fid}}_{\mathrm{lensed}}(z_s)$ to simulate the observational data, resulting in the observed magnitudes $m^{\mathrm{obs}}_{\mathrm{lensed}}(z_s)$. The results of simulated lensed magnitude observations are presented in Figure~\ref{Fig:mzlensed}.

% \subsection{Strong Gravitational Lensing estimations }
% Under the assumption of a power-law lens model with four lensed images, we employ \texttt{lenstronomy} \citep{Birrer:2021use} to simulate two critical observables for a gravitationally lensed SNe Ia system within the context of the fiducial cosmological model: the time-delay distance $D_{\Delta t}^{\rm fid}$ and the angular diameter distance to the lens galaxy $D_d^{\rm fid}$. Moreover, we add a intrinsic mass-sheet parameter $\lambda = 1.2$ to the results from fiducial cosmological model following
% \begin{equation}
%     D_{\Delta t}^{\rm PL} = D_{\Delta t}^{\rm fid}/\lambda
% \end{equation}
% and 
% \begin{equation}
%     D_{d}^{\rm PL} = D_{d}^{\rm fid}
% \end{equation}
% The final estimation of these distance parameters are shown in Figure~\ref{Fig:ddt}.

% The corresponding magnification factors for the multiple lensed images are presented in Figure~\ref{Fig:mag}.
% \begin{figure}
% \centering
% \includegraphics[width=0.5\textwidth]{Lens01-Ddt_ddPL.pdf}
% \caption{The estimated time-delay distance and the angular-diameter distance to the lens based on a power-law lens model with $z_{\rm{s}} = 1.5$, $z_{\rm{d}}=0.5$.}
%   \label{Fig:ddt}%
% \end{figure}

% \begin{figure}
% \centering
% \includegraphics[width=0.5\textwidth]{Lens01-mu4PL.pdf}
% \caption{The estimated magnifications for the four images based on a power-law lens model with $z_{\rm{s}} = 1.5$, $z_{\rm{d}}=0.5$.}
%   \label{Fig:mag}%
% \end{figure}

\section{Methodology and Results} \label{sec:method}
In this section, we delineate the methodology for addressing the MSD utilizing both lensed and unlensed SN Ia observations.  
{During our data analysis, we use a Python package named \texttt{emcee} \citep{Foreman-Mackey:2012any} to do the Markov Chain Monte Carlo (MCMC) analysis, and we analyze the final results with \texttt{GetDist} \citep{Lewis:2019xzd}.
}

\subsection{Breaking the MSD with lensing magnifications} 
We further elucidate the Bayesian framework devised to constrain the intrinsic MSD parameter $\lambda$ through the analysis of gravitational lensing magnification measurements.

\begin{figure}
\centering
\includegraphics[width=0.5\textwidth]{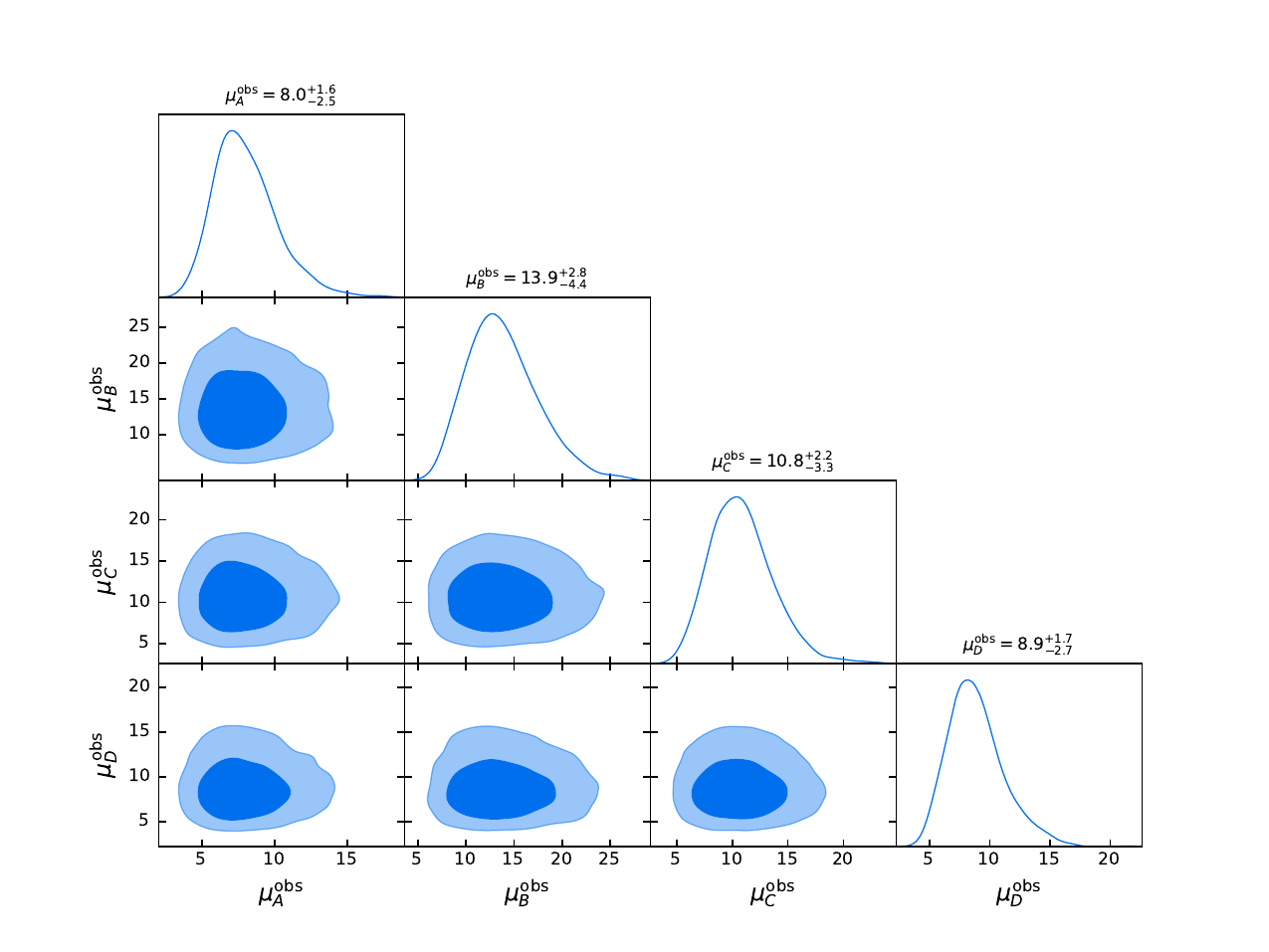}
\caption{Simulated observational magnifications for lensed Type Ia supernovae at $z_{\rm{s}} = 1.5$, $z_{\rm{d}}=0.5$.}
  \label{Fig:muobs}%
\end{figure}

For each lensed image $i \in {A,B,C,D}$, we calculate the observed magnification factor using photometric measurements from the lensed system and GP extrapolations of the unlensed source brightness. The magnification is derived as:
\begin{equation} \label{eq:muobs}
\mu_i^{\text{obs}}(z_s) = 10^{\left[ -\frac{m^{\mathrm{obs}}_{\mathrm{lensed}}(z_s)-m^{\mathrm{GP}}_{\mathrm{unlensed}}(z_s)}{2.5}  \right]}
\end{equation}
where $m^{\mathrm{obs}}_{\mathrm{lensed}}(z_s)$ is the observations of the lensed magnitudes for each image (see figure~\ref{Fig:mzlensed}) and 
$m^{\mathrm{GP}}_{\mathrm{unlensed}}(z_s)$ represents 1,000 independent realizations of the unlensed magnitude posterior from our GP analysis, which are shown in figure~\ref{Fig:mzunlensed}.
Then, we can obtain the distribution of $\mu_i^{\text{obs}}$, which is illustrated in figure~\ref{Fig:muobs}.

The theoretical magnification under the power-law lens model is related to the intrinsic mass sheet parameter through
\begin{equation}\label{eq:muth}
   \mu^{{\rm{th}}}_i = \mu^{\rm{PL}}_i/\lambda^2 
\end{equation}
where $\mu_i^{\mathrm{PL}}$ denotes the magnification estimated by the baseline power-law lens model.

Finally, we construct the joint posterior distribution through likelihood multiplication and prior incorporation:
\begin{equation}
\label{eq:posterior01}
    P(\lambda,\mu_i^{\rm{PL}} | \mu_i^{\text{obs}}) \propto P(\mu_i^{\text{obs}}|\lambda,\mu_i^{\rm{PL}})P(\mu_i^{\rm{PL}})P(\lambda),
\end{equation}
in which a flat prior $\lambda \in [0.5, 1.5]$ is assumed and the priors for $\mu_i^{\rm{PL}}$ are provided in Figure~\ref{Fig:muiddt}. 
{Here we treat $\mu^{\rm{PL}}_i$ and $\lambda$ as free parameters to be estimated and the magnifications $\mu^{\rm{PL}}_i$ estimated by the baseline power-law lens model from \textsc{Lenstronomy} as prior.}

We impose constraints on the MSD parameter to $\lambda = 1.191_{-0.110}^{+0.094}$, maintaining a credibility interval of $68\%$, which demonstrates strong statistical alignment with the fiducial value $\lambda^{\mathrm{fid}} = 1.2$. This concurrence corroborates the methodology employed in this study. Nevertheless, we observe degeneracies between $\lambda$ and the individual magnification parameters $\mu_i^{\rm{PL}}$.

\begin{figure*}
\centering
\includegraphics[width=0.75\textwidth]{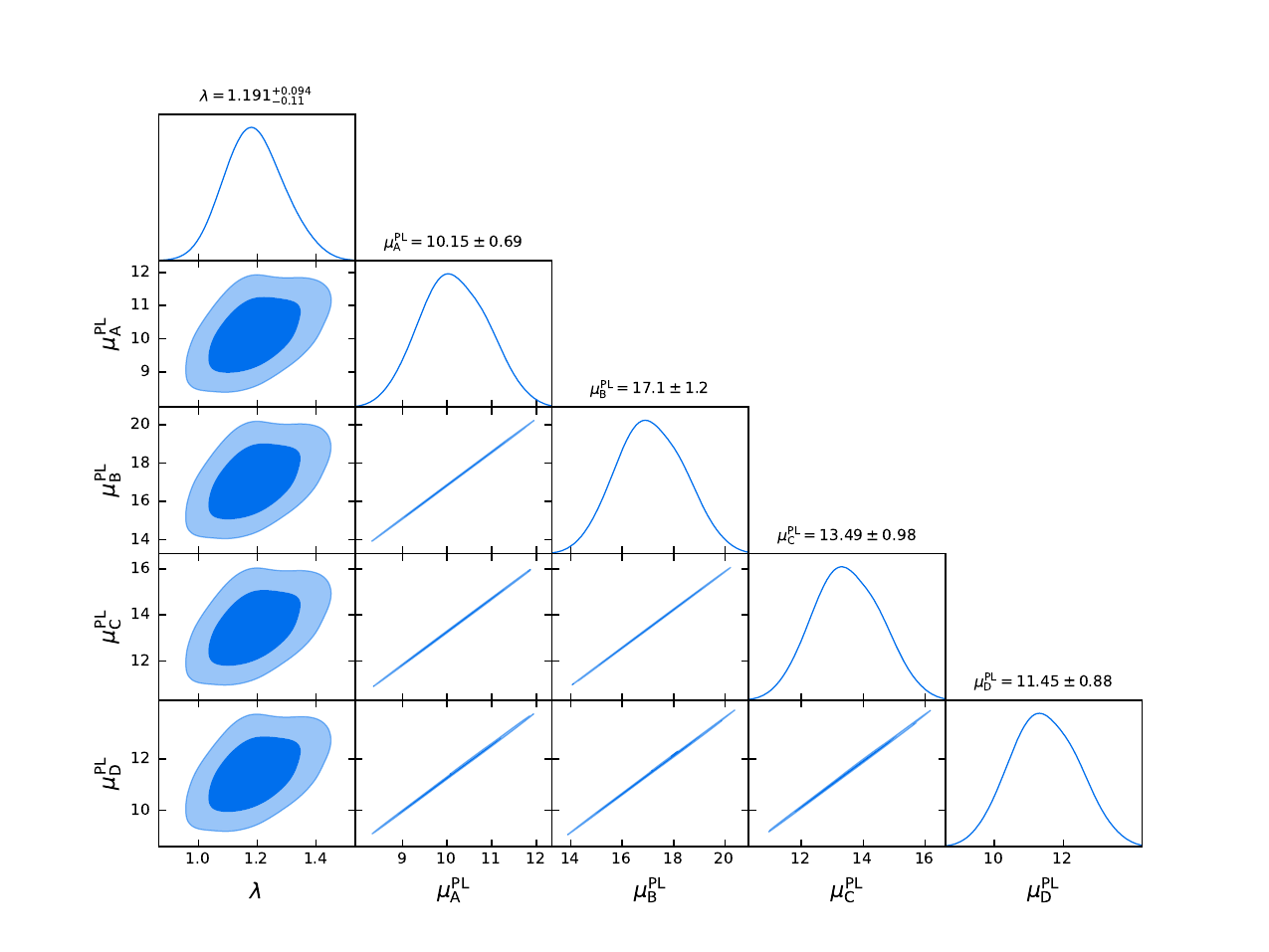}
\caption{The posterior results on $\lambda$ and $\mu_i^{\rm{PL}}$ obtained from Eq.~(\ref{eq:posterior01}) for the simulated lensed SN Ia at $z_d = 0.5$ and $z_s = 1.5$ which have four images.}
  \label{Fig:res_lambdamui}%
\end{figure*}

\subsection{Breaking the MSD with distance ratio}
We present a geometric approach to break the MSD using cosmological distance measurements in this subsection. 

In a flat universe, we can convey $(D_{\rm{L}}H_0(z))^{\rm{GP}}$ to $(D_{\rm{A}}H_0(z))^{\rm{GP}}$ with
\begin{equation}
    (D_{\rm{L}}H_0(z))^{\rm{GP}}=\frac{(D_{\rm{A}}H_0(z))^{\rm{GP}}}{(1+z)^2}.
\end{equation}

\begin{figure}
\centering
\includegraphics[width=0.45\textwidth]{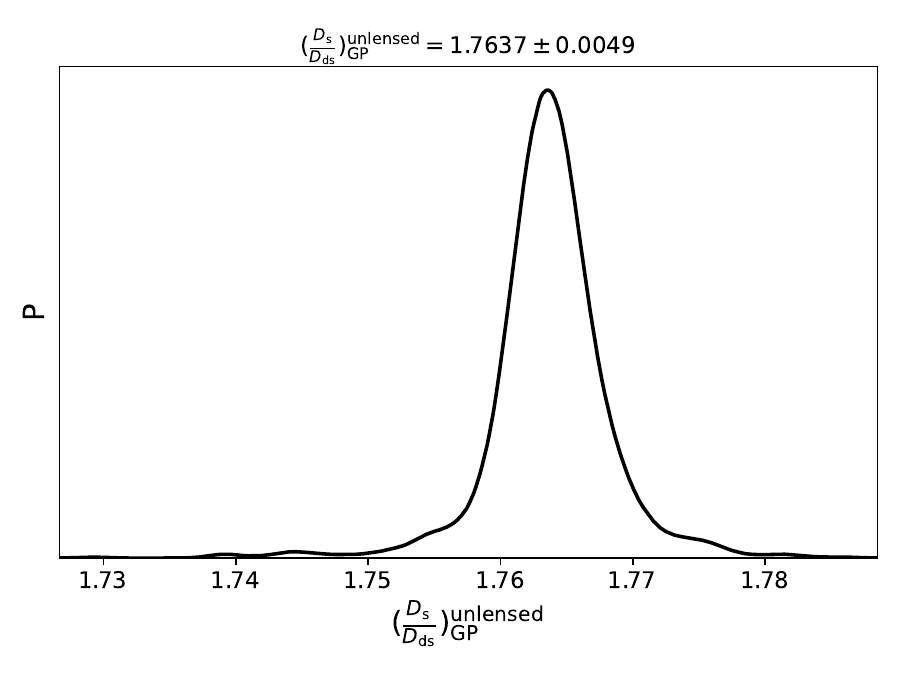}
\caption{The simulated observations of distance ratio for lensed SN Ia with $z_{\rm{s}} = 1.5$, $z_{\rm{d}}=0.5$.}
  \label{Fig:DDobs}%
\end{figure}

And then we can have the unanchored angular diameter distance between the source and the lens $(D_{\rm{ds}}H_0)^{\rm{GP}}$ following:
\begin{equation}
(D_{\rm{ds}}H_0)^{\rm{GP}} =(D_{\rm{s}}H_0)^{\rm{GP}}-(D_{\rm{d}}H_0)^{\rm{GP}} \frac{1+z_d}{1+z_s}   
\end{equation}
where $D_{\rm{s}} = D_A(z_s)$ is the angular diameter distance to the background source and $D_{\rm{d}} = D_A(z_d)$ is the angular diameter distance to the lens.
 The distance ratio $(\frac{D_{\rm{s}}}{D_{\rm{ds}}})^{\rm{unlensed}}_{\rm{GP}}$ can be obtained with
\begin{equation}\label{eq:DsDdsobs}
    (\frac{D_{\rm{s}}}{D_{\rm{ds}}})^{\rm{unlensed}}_{\rm{GP}} = \frac{(D_{\rm{s}}H_0)^{\rm{GP}}}{(D_{\rm{ds}}H_0)^{\rm{GP}}}
\end{equation}
We show the distribution of $(\frac{D_{\rm{s}}}{D_{\rm{ds}}})^{\rm{unlensed}}_{\rm{GP}}$ in Figure~\ref{Fig:DDobs}.

The theoretical distance ratio under the power-law lens model is related to the MSD parameter through
\begin{equation}\label{eq:DDth}
(\frac{D_{\rm{s}}}{D_{\rm{ds}}})^{\rm{th}} = (\frac{D_{\rm{s}}}{D_{\rm{ds}}})^{\rm{PL}}/\lambda
\end{equation}
where $(\frac{D_{\rm{s}}}{D_{\rm{ds}}})^{\mathrm{PL}}$ denotes the distance ratio predicted by the baseline power-law lens model, which can be obtained through
\begin{equation}
    (\frac{D_{\rm{s}}}{D_{\rm{ds}}})^{\rm{PL}} = \frac{D_{\Delta t}^{\rm{PL}}}{(1+z_{\rm{d}})D_{\rm{d}}^{\rm{PL}}}
\end{equation}

By using the simulated observational dataset of $(\frac{D_{\rm{s}}}{D_{\rm{ds}}})^{\rm{unlensed}}_{\rm{GP}}$ combined with the estimated time-delay distance $D_{\Delta t}^{\rm{PL}}$ and lens angular diameter distance $D_{\rm{d}}^{\rm{PL}}$, we perform Bayesian posterior inference to simultaneously constrain the parameter $\lambda$ and the poeterior of both $D_{\Delta t}^{\rm{PL}}$ and $D_{\rm{d}}^{\rm{PL}}$, following
\begin{equation}\label{eq:lamDsDdt}
\begin{aligned}
   & P(\lambda,D_{\Delta t}^{\rm{PL}},D_{\rm{d}}^{\rm{PL}} | (\frac{D_{\rm{s}}}{D_{\rm{ds}}})^{\rm{unlensed}}_{\rm{GP}}) \propto \\
     &P((\frac{D_{\rm{s}}}{D_{\rm{ds}}})^{\rm{unlensed}}_{\rm{GP}}|\lambda,D_{\Delta t}^{\rm{PL}},D_{\rm{d}}^{\rm{PL}})P(D_{\Delta t}^{\rm{PL}},D_{\rm{d}}^{\rm{PL}})P(\lambda),
\end{aligned}
\end{equation}
where a flat prior $\lambda \in [0.5, 1.5]$ is assumed and the priors for $(D_{\Delta t}^{\rm{PL}},D_{\rm{d}}^{\rm{PL}})$ are provided in Figure~\ref{Fig:muiddt}.
The distribution of $(\frac{D_{\rm{s}}}{D_{\rm{ds}}})^{\rm{unlensed}}_{\rm{GP}}$ are shown in figure~\ref{Fig:DDobs}.
We treat $D_{\Delta t}^{\rm{PL}},D_{\rm{d}}^{\rm{PL}}$ and $\lambda$ as free parameters to be estimated.

\begin{figure}
\centering
\includegraphics[width=0.45\textwidth]{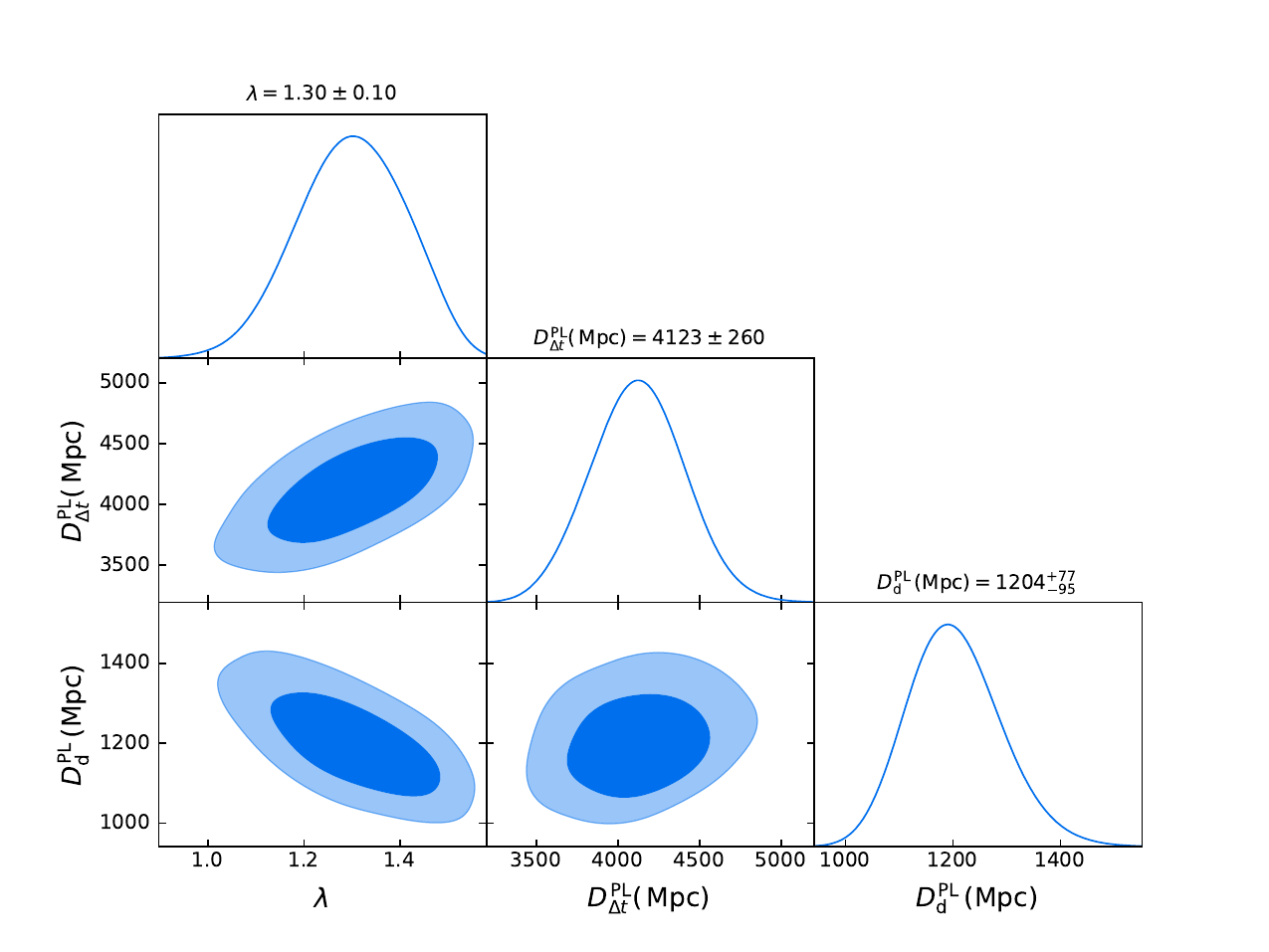}
\caption{The posterior results on $\lambda$ and $(\frac{D_{\rm{s}}}{D_{\rm{ds}}})^{\rm{PL}}$ obtained from Eq.~(\ref{eq:lamDsDdt}) for the simulated lensed SN Ia at $z_d = 0.5$ and $z_s = 1.5$ which have four images.}
  \label{Fig:res_lambdaDD}%
\end{figure}

By rigorously applying the aforementioned robust methodology, we successfully constrained the MSD parameter to 
$\lambda = 1.30 \pm 0.10$ ($68\%$ CL), which is in good agreement with the fiducial value 
$\lambda^{\mathrm{fid}} = 1.2$. Figure~\ref{Fig:res_lambdaDD} shows the joint posterior distributions of 
$\lambda$ and ($D_{\Delta t}^{\mathrm{PL}}, D_d^{\mathrm{PL}}$), illustrating our comprehensive analysis. Notably, the posterior distribution reveals a strong degeneracy between 
$\lambda$ and $D_{\Delta t}^{\mathrm{PL}}, D_d^{\mathrm{PL}}$, suggesting further investigation into the underlying dynamics.

\subsection{Breaking the MSD with the combinations of lensing magnifications and distance ratio}
\begin{figure*}
\centering
\includegraphics[width=0.75\textwidth]{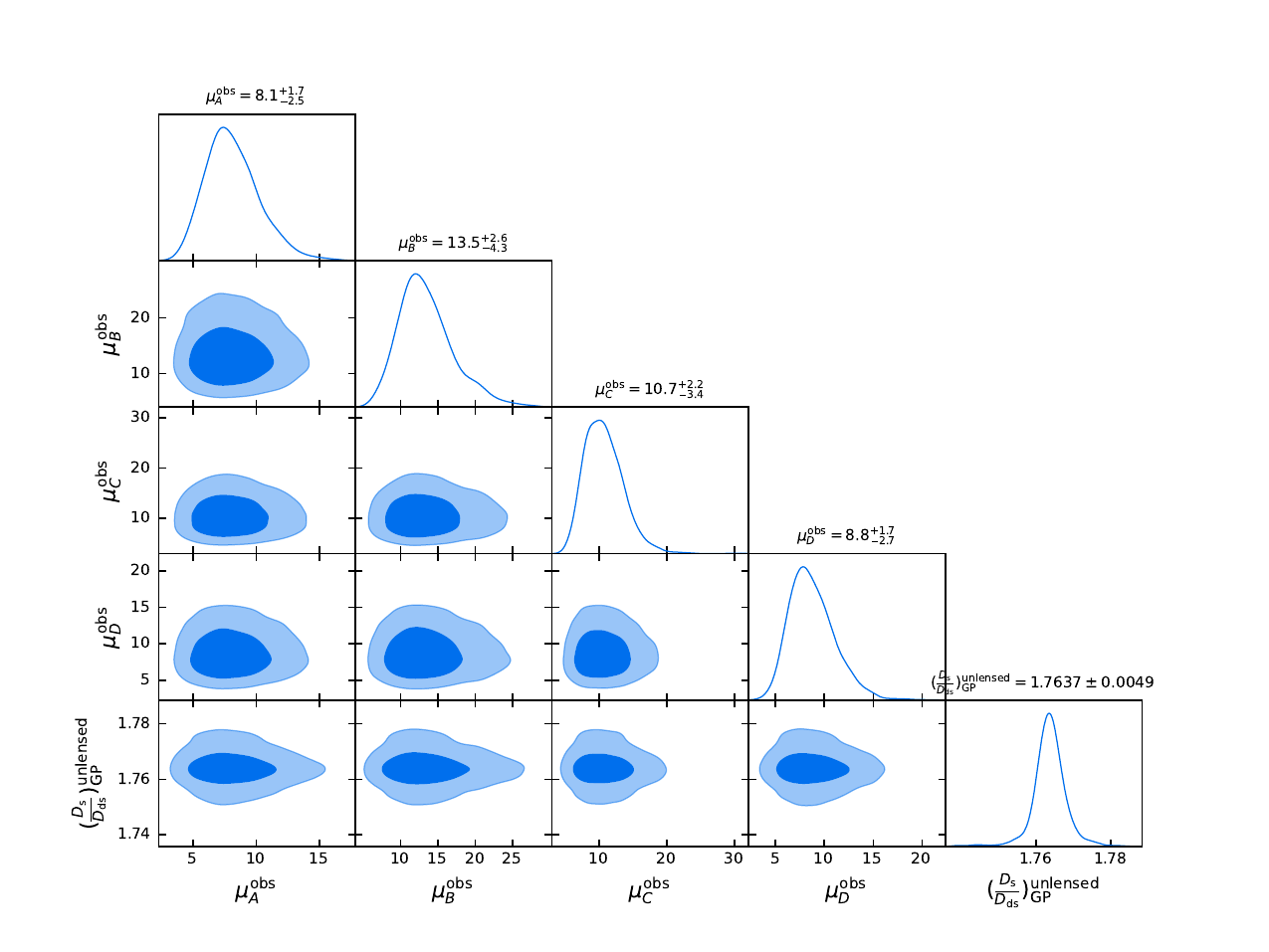}
\caption{The joint distribution of the distance ratio $(D_s/D_ds)^{\rm unlensed}_{\rm GP}$ (Eq.~\ref{eq:DsDdsobs}) and the magnifications $\mu_i^{\rm obs}$ (Eq.~\ref{eq:muobs}), constructed from 1,000 Gaussian process realizations of the unlensed SNe Ia combined with the simulated lensed SN Ia data ($z_s=1.5$, $z_d=0.5$).}
  \label{Fig:muiDDobs}%
\end{figure*}

\begin{figure*}
\centering
\includegraphics[width=0.85\textwidth]{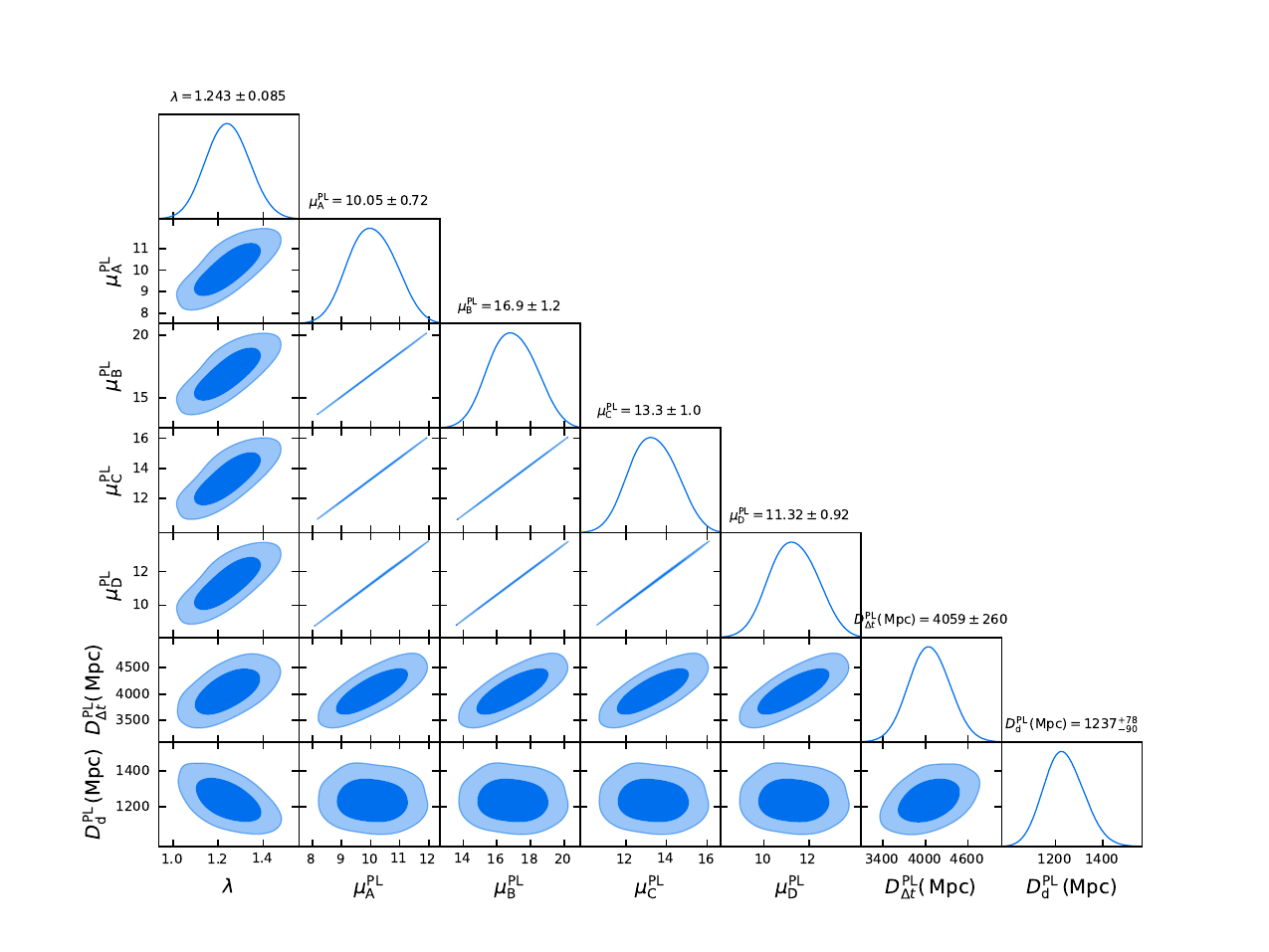}
\caption{The joint posterior inference of the MST parameter $\lambda$ and the power-law modeled magnifications $\mu_i^{\rm{PL}}$ as well as the distance ratio $(\frac{D_{\rm{s}}}{D_{\rm{ds}}})^{\rm{PL}}$ obtained from Eq. (\ref{eq:posterior_com}).}
  \label{Fig:lambdaintmuiDsDdt}%
\end{figure*}

By employing the aforementioned methodology, the MSD can be resolved through the combinations of gravitational lensing magnifications and cosmic distance ratios. The joint posterior inference of $\lambda$, $\mu_i^{\rm{PL}}$, and $(\frac{D_{\rm{s}}}{D_{\rm{ds}}})^{\rm{PL}}$ can be obtained through
\begin{equation}\label{eq:posterior_com}
\begin{aligned}
   & P\left(\lambda,\mu_i^{\rm{PL}},D_{\Delta t}^{\rm{PL}},D_{\rm{d}}^{\rm{PL}} | (\frac{D_{\rm{s}}}{D_{\rm{ds}}})^{\rm{unlensed}}_{\rm{GP}},\mu_i^{\rm{obs}} \right) \propto  \\
     &P\left((\frac{D_{\rm{s}}}{D_{\rm{ds}}})^{\rm{unlensed}}_{\rm{GP}},\mu_i^{\rm{obs}}|\lambda,D_{\Delta t}^{\rm{PL}},D_{\rm{d}}^{\rm{PL}},\mu_i^{\rm{PL}} \right)\times  \\
      &P\left(D_{\Delta t}^{\rm{PL}},D_{\rm{d}}^{\rm{PL}},\mu_i^{\rm{PL}} \right)  P(\lambda),
\end{aligned}
\end{equation}
    where the joint distribution of the distance ratio $(\frac{D_{\rm{s}}}{D_{\rm{ds}}})^{\rm{unlensed}}{\rm{GP}}$ and the magnifications $\mu_i^{\rm{obs}}$ are shown in Figure~\ref{Fig:muiDDobs}.
    This joint distribution is constructed from the 1,000 GP realizations described in Section~\ref{sec:data}. Each GP realization provides both $(D_s/D_{\rm{ds}})^{\rm unlensed}_{\rm GP}$ (via Eq.~\ref{eq:DsDdsobs}) and, when combined with the observed magnitudes of the four images, the magnifications $\mu_i^{\rm obs}$ (via Eq.~\ref{eq:muobs}).
To evaluate the likelihood term $P\left((D_s/D_{ds})^{\rm unlensed}_{\rm GP},\mu_i^{\rm obs} \mid \lambda, D_{\Delta t}^{\rm PL}, D_d^{\rm PL}, \mu_i^{\rm PL}\right)$ during MCMC, we first calculate the predicted values $(D_s/D_{ds})^{\rm th}$ and $\mu_i^{\rm th}$ from the proposed model parameters via Eqs.~(\ref{eq:DDth}) and~(\ref{eq:muth}).
We then assess the probability density of these predictions under the observed joint distribution. For this purpose we employ a Gaussian kernel density estimator (KDE) fitted to the 1,000 samples of $((D_s/D_ds)^{\rm unlensed}_{\rm GP},\mu_i^{\rm obs})$.
The KDE provides a smooth, non‑parametric representation of the underlying probability density, accounting for correlations and non‑Gaussian features.
At each MCMC step, the likelihood is taken as the KDE value at the predicted point. This procedure ensures that the full uncertainty information from the GP reconstruction is propagated into the final parameter constraints and it is analogous to the use of posterior samples in hierarchical Bayesian analyses such as TDCOSMO IV \citep{Birrer:2020tax}, where individual lens constraints are marginalized over population distributions.

{From Figure~\ref{Fig:muiDDobs}, we found $\mu_i^{\rm obs}$ are very wealy correlated with $(\frac{D_{\rm{s}}}{D_{\rm{ds}}})^{\rm{unlensed}}_{\rm{GP}}$, besides, $\mu_i^{\rm obs}$ among the 4 images are nearly independent. The reasons could be: $(\frac{D_{\rm{s}}}{D_{\rm{ds}}})^{\rm{unlensed}}_{\rm{GP}}$ is got with the measurements at $z_{\rm d}$ and $z_{\rm s}$ with unlensed SNe Ia, while $\mu_i^{\rm obs}$ is got with only $z_{\rm s}$ plus independent measurement of the image brightness. Image brightnesses are independently measured among the 4 images (photometry) even though they share a comment unlensed brightness from unlensed SNe.}
The prior for $D_{\Delta t}^{\rm{PL}},D_{\rm{d}}^{\rm{PL}}$ and $\mu_i^{\rm{PL}}$ are shown in figure~\ref{Fig:muiddt}. A flat prior $\lambda \in [0.5, 1.5]$ is assumed.

Using a Bayesian framework combining gravitational lensing magnification and cosmic distance ratios, we constrain the MSD parameter to $\lambda = 1.243\pm0.085$ ($6.7\%$ precision), in excellent agreement with the fiducial $\lambda^{\mathrm{fid}} = 1.2$. This represents significant improvement over individual methods: distance ratios yield $\lambda = 1.300\pm0.100$ ($7.6\%$) and magnifications give $\lambda = 1.191_{-0.110}^{+0.094}$ ($8.5\%$). The combined analysis resolves geometric degeneracies in mass distribution while achieving superior precision, validating our methodology. The joint posterior distributions of $\lambda$ and $D_{\Delta t}^{\rm{PL}},D_{\rm{d}}^{\rm{PL}}$ as well as $\mu_i^{\rm{PL}}$  are shown in figure~\ref{Fig:lambdaintmuiDsDdt}. 
% Future work incorporating stellar kinematics and multi-wavelength data could further enhance these constraints.

% \begin{figure*}
% \centering
% \includegraphics[width=0.75\textwidth]{Lens01-mu4_DD_obs.pdf}
% \caption{The joint distribution of the distance ratio $(\frac{D_{\rm{s}}}{D_{\rm{ds}}})^{\rm{unlensed}}_{\rm{GP}}$ and the magnifications $\mu_i^{\rm{obs}}$  for lensed SN Ia with $z_{\rm{s}} = 1.5$, $z_{\rm{d}}=0.5$.}
%   \label{Fig:muiDDobs}%
% \end{figure*}

% \begin{figure*}
% \centering
% \includegraphics[width=0.85\textwidth]{Lens01-lambdamu4DD01.pdf}
% \caption{The joint posterior inference of the MST parameter $\lambda$ and the power-law modeled magnifications $\mu_i^{\rm{PL}}$ as well as the distance ratio $(\frac{D_{\rm{s}}}{D_{\rm{ds}}})^{\rm{PL}}$.}
%   \label{Fig:lambdaintmuiDsDdt}%
% \end{figure*}

\section{Framework with multiple systems} \label{sec:multi-systems}
\subsection{Strong Gravitational Lensing estimations }
\begin{figure*}
\centering
\includegraphics[width=0.75\textwidth]{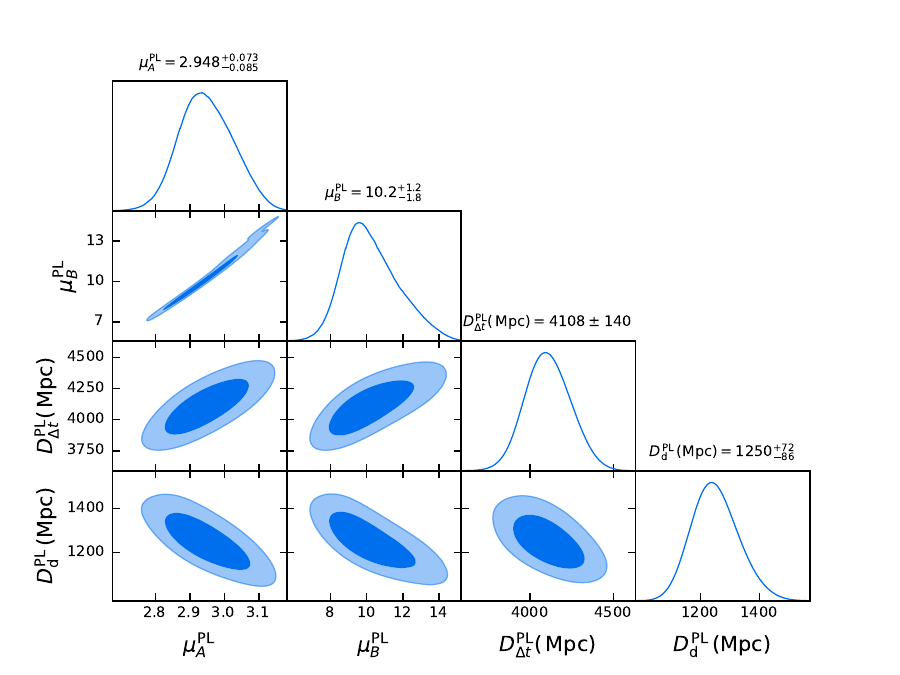}
\caption{The time-delay distance ($D_{\Delta t}$) and angular-diameter distance to the lens ($D_{\rm{d}}$), along with the magnification ratios of the lensed images. The results are derived from a power-law lens model in \textsc{Lenstronomy}, with the incorporation of an intrinsic MSD parameter.}
  \label{Fig:muiddt02}%
\end{figure*}

\begin{figure*}
\centering
\includegraphics[width=1.05\textwidth]{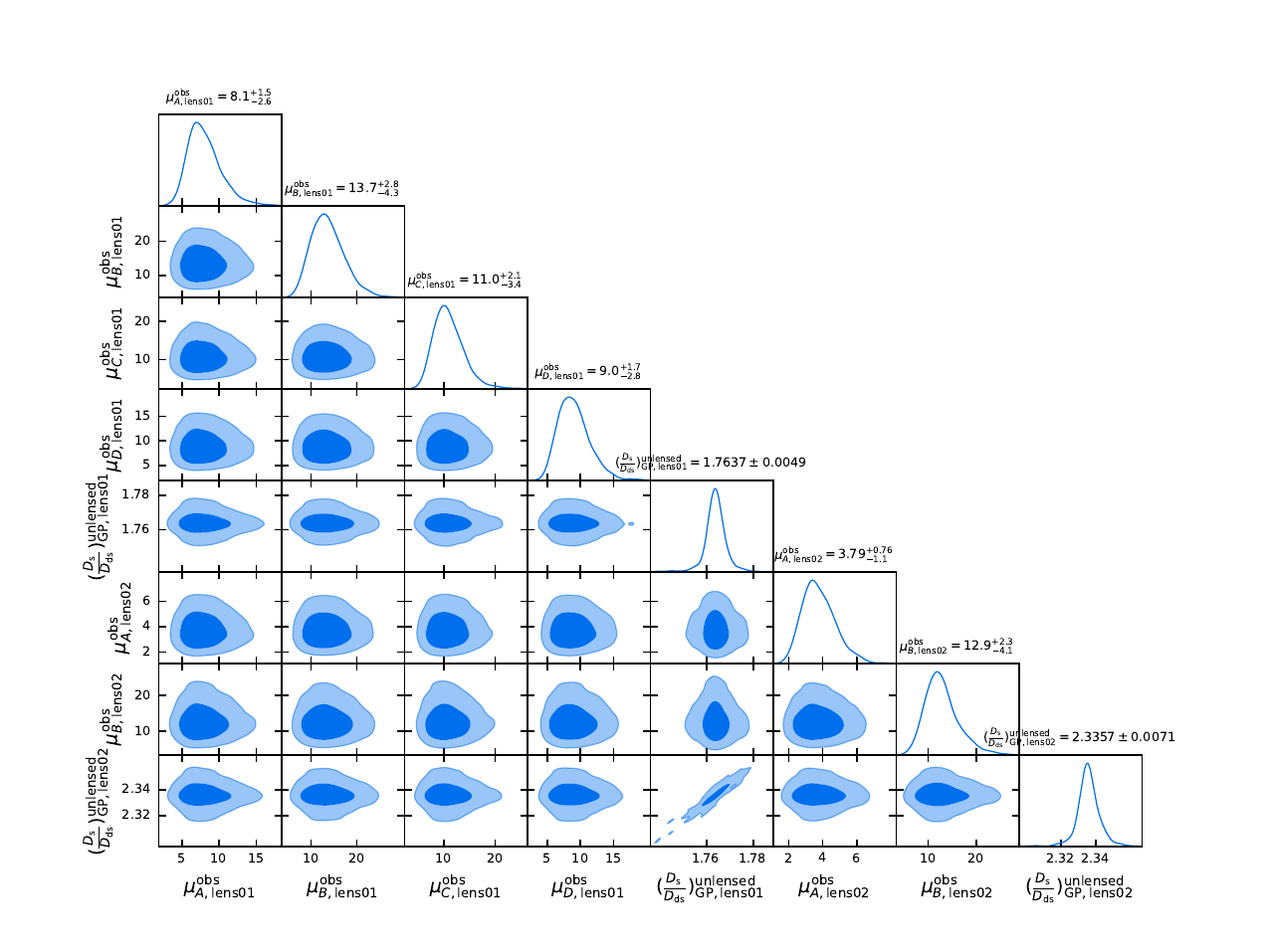}
\caption{The joint priors for magnification factors and the distance ratio for two lensing systems.}
\label{Fig:lens0102muiDsDdt_obs_2lens}
\end{figure*}

Building upon the framework established in Section~\ref{sec:method}, we extend our investigation to address the MSD in two different strong gravitational lensing systems. Under the fiducial cosmology, we generate another lensed SNe Ia which have two images for systems with source redshift at $z_s = 1.0$ and lens redshifts at $ z_d = 0.5$ (We refer to this lensing system as lens 02 hereafter) and following the method described above to do constraints. The main methods are summarized as follows:

\begin{enumerate}
\item Using \textsc{Lenstronomy}, we estimate the power-law modeled magnification factors $\mu_{j,02}^{\rm{PL}}$ (where $j = \{\rm{A,B}\}$ denotes the two lensed images) along with the derived cosmological distances for the lens system 02. The complete observational configuration and system parameters for lens 02 are provided in Column 4 of Table~\ref{tab:lenstronomy}. The final estimation of the magnification factors $\mu_{j,02}^{\rm{PL}}$ and the distances $(D_{\Delta t,02}^{\rm{PL}},D_{\rm{d,02}}^{\rm{PL}})$ are shown in Figure~\ref{Fig:muiddt02}.

  \item Simulate the observations of magnitude $m^{\rm{obs}}_{\rm{lensed, 02}}(z_{s,02})$ and distance ratio $(\frac{D_{\rm{s}}}{D_{\rm{ds}}})_{02}^{\rm{obs}}$ for lensed 02 based on the fiducial cosmological model.
  
\item Derive the joint distribution of $\mu_{i,01}^{\rm{obs}}$, $(\frac{D_{\rm{s}}}{D_{\rm{ds}}})_{\rm GP, 01}^{\rm{unlensed}}$, $\mu_{j,02}^{\rm{obs}}$ and $(\frac{D_{\rm{s}}}{D_{\rm{ds}}})_{\rm GP, 02}^{\rm{unlensed}}$ from the observed data $m^{\rm{obs}}_{\rm{lensed, 01}}(z_{s,01})$, $m^{\rm{obs}}_{\rm{lensed, 02}}(z_{s,02})$, $(\frac{D_{\rm{s}}}{D_{\rm{ds}}})_{01}^{\rm{obs}}$, and $(\frac{D_{\rm{s}}}{D_{\rm{ds}}})_{02}^{\rm{obs}}$, in conjunction with GP realizations of $m_{i,01}^{\rm{GP}}$, $m_{j,02}^{\rm{GP}}$, $(\frac{D_{\rm{s}}}{D_{\rm{ds}}})_{01}^{\rm{GP}}$, and $(\frac{D_{\rm{s}}}{D_{\rm{ds}}})_{02}^{\rm{GP}}$, according to equations~(\ref{eq:muobs}) and (\ref{eq:DsDdsobs}). Here, $i = A, B, C, D$ represent the four images associated with lens 01, while $j = A, B$ correspond to the two images pertaining to lens 02. The joint distribution of the magnifications, time-delay distance and the angular diameter distance to the lens for the two lenses are shown in Figure~\ref{Fig:lens0102muiDsDdt_obs_2lens}.
\item Calculate the joint posterior inference of the intrinsic MSD parameter $\lambda$ as well as the power-law modeled magnifications and the distances for the two lensing systems with

\begin{widetext}
\begin{equation}\label{eq:lamDsDdt2lens}
\begin{aligned}
   & P\left(\lambda_{\rm{01}},\lambda_{\rm{02}},\mu_{i,01}^{\rm{PL}},\mu_{j,02}^{\rm{PL}},D_{\Delta t,01}^{\rm{PL}},D_{\rm{d},01}^{\rm{PL}},D_{\Delta t,02}^{\rm{PL}},D_{\rm{d},02}^{\rm{PL}} | \mu_{i,01}^{\rm{obs}},(\frac{D_{\rm{s}}}{D_{\rm{ds}}})_{\rm GP, 01}^{\rm{unlensed}},\mu_{j,02}^{\rm{obs}},(\frac{D_{\rm{s}}}{D_{\rm{ds}}})_{\rm GP, 02}^{\rm{unlensed}} \right) \propto \\
     &P\left(\mu_{i,01}^{\rm{obs}},(\frac{D_{\rm{s}}}{D_{\rm{ds}}})_{\rm GP, 01}^{\rm{unlensed}},\mu_{j,02}^{\rm{obs}},(\frac{D_{\rm{s}}}{D_{\rm{ds}}})_{\rm GP, 02}^{\rm{unlensed}}|\lambda_{\rm{01}},\lambda_{\rm{02}},\mu_{i,01}^{\rm{PL}},\mu_{j,02}^{\rm{PL}},D_{\Delta t,01}^{\rm{PL}},D_{\rm{d},01}^{\rm{PL}},D_{\Delta t,02}^{\rm{PL}},D_{\rm{d},02}^{\rm{PL}} \right)\times  \\
      &P\left(D_{\Delta t,01}^{\rm{PL}},D_{\rm{d},01}^{\rm{PL}},\mu_{i,01}^{\rm{PL}} \right)  P(\lambda_{\rm{01}})P\left(D_{\Delta t,02}^{\rm{PL}},D_{\rm{d},02}^{\rm{PL}},\mu_{j,02}^{\rm{PL}} \right)  P(\lambda_{\rm{02}}).
\end{aligned}
\end{equation}
\end{widetext}

\end{enumerate}

% The results are shown in figure~\ref{Fig:lambdaintmuiDD2lens}. 
\begin{figure*}
\centering
\includegraphics[width=0.75\textwidth]{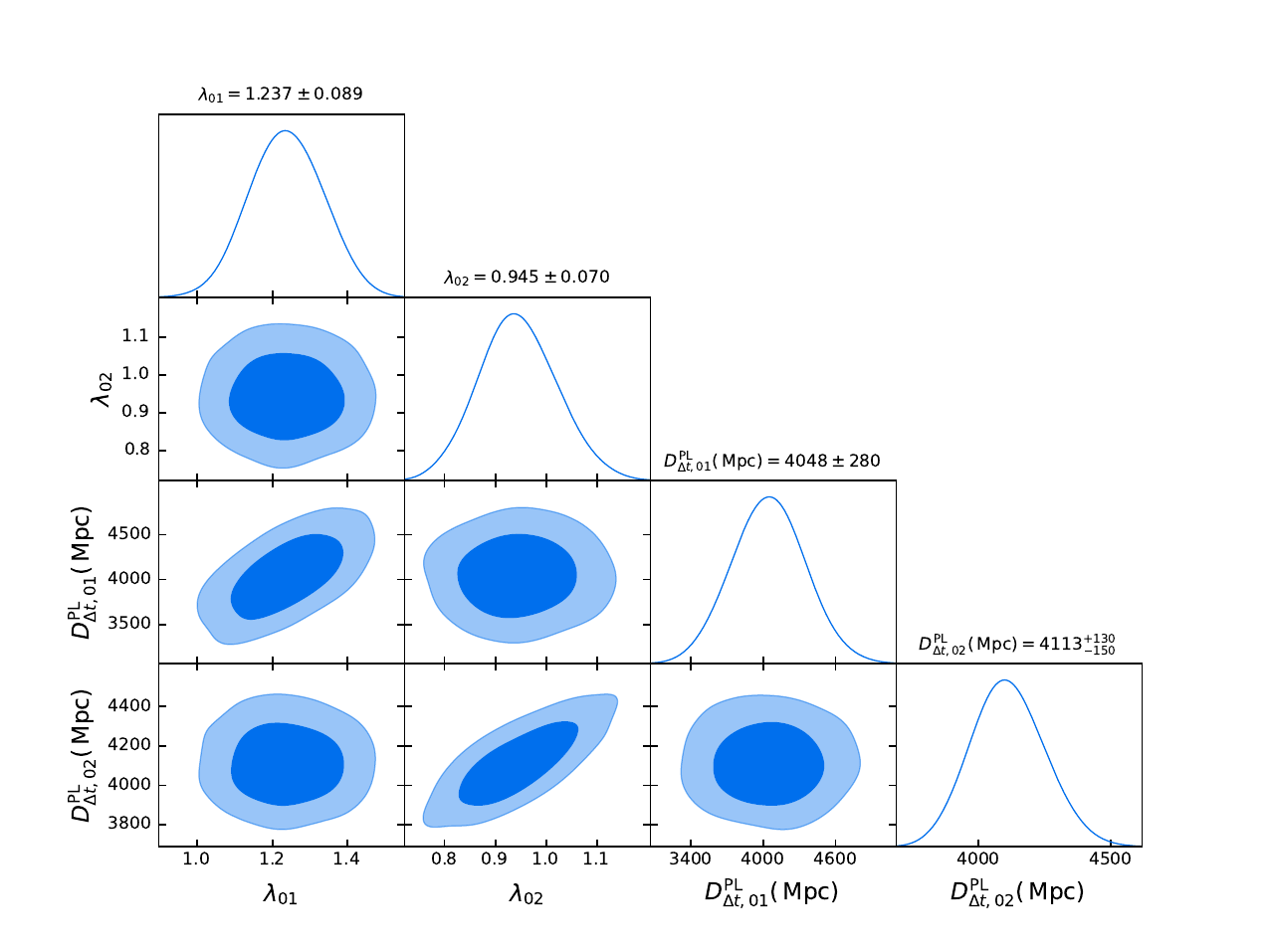}
\caption{Joint post probability distributions for the MST parameters as well as the power-law modeled magnification factors ($\mu^{\rm{PL}}$) and the time-delay distance  ($D_{\Delta{t}}^{\rm{PL}}$) for the two strong lensing systems (Lens 01 and Lens 02) obtained from Eq.~(\ref{eq:lamDsDdt2lens}). }
\label{Fig:lambdaintmuiDD2lens}
\end{figure*}

We have derived the MSD parameters for the two lens systems: $\lambda_{\rm 01} = 1.237\pm 0.089$ pertains to lens 01 and $\lambda_{\rm 02} = 0.945\pm{0.070}$ pertains to lens 02. The measured values show excellent agreement with the fiducial MSD parameters implemented in our simulation framework. This concordance demonstrates the efficacy of our method in breaking the MSD through the combination of lensing magnifications and distance ratios, effectively addressing this fundamental degeneracy. The joint constraints and posterior distributions of these parameters, including their uncertainties and correlations, are visualized in the confidence contours shown in Figure~\ref{Fig:lambdaintmuiDD2lens}.

\section{The measurement of time-delay distance under MST}\label{sec:time-delayDistance}

\begin{figure*}
\centering
\includegraphics[width=0.75\textwidth]{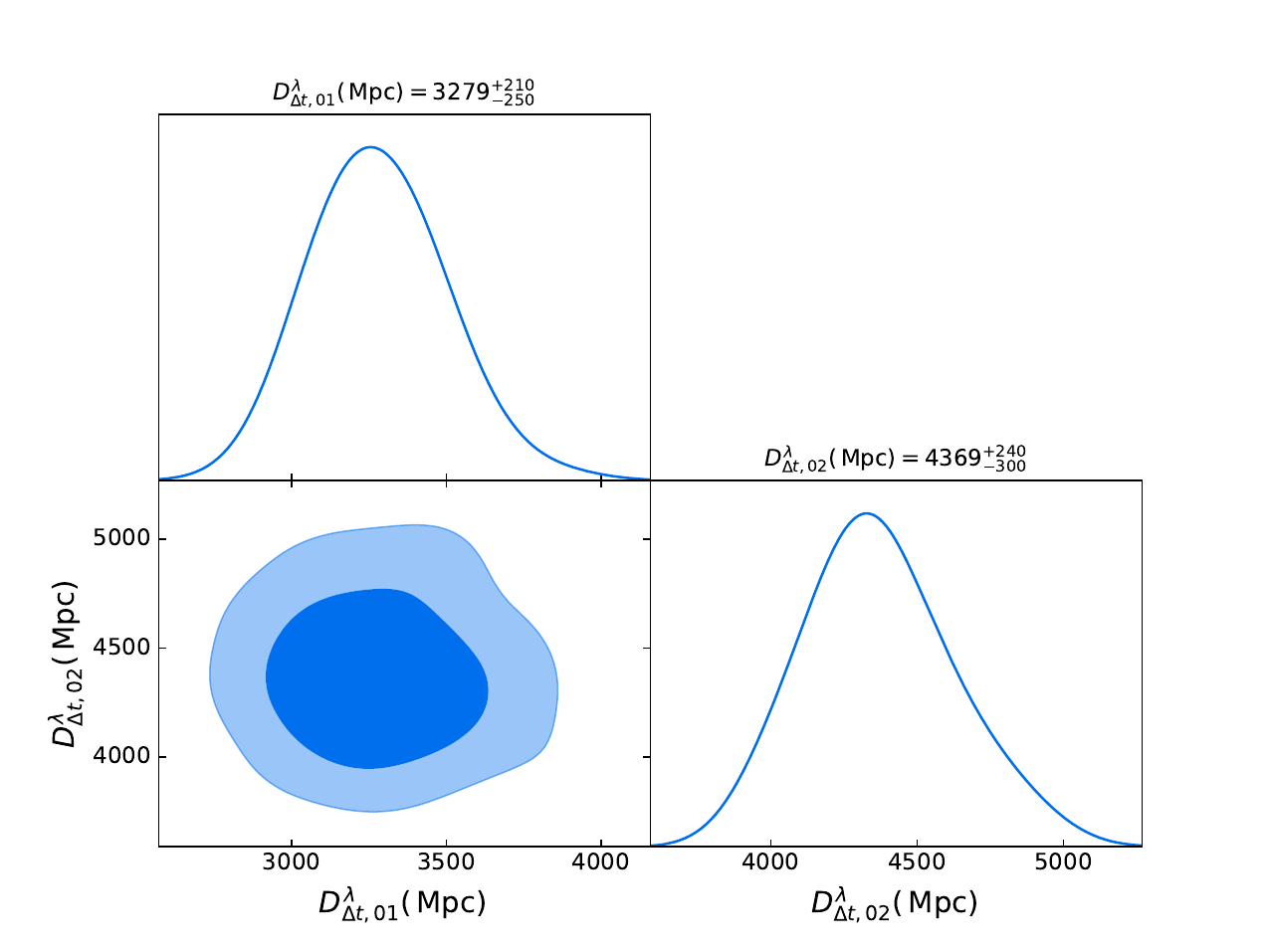}
\caption{The posterior inference  for the corrected time-delay distance considering the MST parameters of the two gravitational lenses.}
\label{Fig:Dt01Dt02}
\end{figure*}

The joint posterior inference of the MSD parameters and time-delay distances for both strong lensing systems, as presented in Figure~\ref{Fig:lambdaintmuiDD2lens}, enables the derivation of corrected time-delay distances accounting for the mass-sheet transformation (MST). The transformation from the power-law modeled distances to the physical distances is governed by:

\begin{equation}
    D_{\Delta t}^{\lambda} = \frac{D_{\Delta t}^{\rm PL}}{\lambda},
\end{equation}
where $\lambda$ represents the MSD parameter constrained by our Bayesian analysis. The resulting corrected time-delay distances for both lens systems are displayed in Figure~\ref{Fig:Dt01Dt02}, showing significantly corrections compared to the power-law model estimates.

The correction procedure offers significant advantages for cosmological applications by rectifying inherent biases in power-law mass model assumptions. This transformation maintains all observable constraints while properly accounting for the MSD, with our lens sample showing typical correction factors $\lambda$ ranging from 0.8 to 1.2 (68\% confidence level). These adjustments are crucial for obtaining unbiased estimates of lensing quantities that would otherwise be affected by simplistic power-law approximations.

Most importantly, this methodology enhances the precision of $H_0$ measurements through multiple mechanisms. It reduces model-dependent systematics compared to conventional power-law analyses while simultaneously breaking the degeneracy between mass profile slope and absolute mass normalization. Furthermore, the framework provides robust, self-consistent error propagation from lens models to cosmological parameters, ensuring reliable uncertainty estimates in the final $H_0$ determination. Together, these improvements enable more accurate cosmological constraints from strong lensing systems.

The implementation of this correction framework demonstrates that proper accounting for the MSD in strong lensing analyses is essential for achieving percent-level precision in $H_0$ measurements, particularly when combining multiple lens systems with varying image configurations and redshift distributions.

% With the posterior inference for the MSD parameters and the time-delay distances of the two strong gravitational lens systems shown in Figure~\ref{Fig:lambdaintmuiDD2lens}, one can derive the corrected measurements of the time-delay distance under the MST through
% \begin{equation}
%     D_{\Delta t}^{\lambda} = D_{\Delta t}^{\rm PL}/\lambda.
% \end{equation}
% The results are shown in Figure~\ref{Fig:Dt01Dt02}.

% Upon deriving the corrections to the time-delay distance via the MST parameter $\lambda$, the rectification of the power-law model becomes attainable. This methodology subsequently augments the precision of Hubble constant measurements by efficaciously alleviating model-dependent degeneracies within strong gravitational lensing systems.

% \begin{figure*}
% \centering
% \includegraphics[width=0.95\textwidth]{Lens0102_Dt01Dt02lmDt.pdf}
% \caption{The posterior inference  for the corrected time-delay distance considering the MST parameters of the two gravitational lenses.}
% \label{Fig:Dt01Dt02lmDt}
% \end{figure*}

\section{Conclusion and Discussion} \label{sec:con}

In this work, we present a novel methodology to resolve the persistent MSD in time-delay cosmography through the using of both gravitationally lensed and unlensed SNe Ia. Our key findings and implications can be summarized as follows:

\begin{enumerate}

\item By simultaneously utilizing lensing magnification measurements ($\mu^{\rm obs}$) and cosmic distance ratios ($D_s/D_{ds}$), we establish a Bayesian framework capable of breaking the MSD without prior assumptions about lens mass profiles. The joint likelihood analysis constrains the MSD parameter $\lambda$ to $\lambda = 1.243\pm0.085$ with $6.7\%$ relative precision, demonstrating significant improvement over individual constraint methods (7.6$\%$ from magnifications alone, $8.5\%$ from distance ratios).

\item 
GP of unlensed SNe Ia magnitudes enables non-parametric mapping of luminosity distances $D_L^{\rm GP}(z)$, effectively decoupling MSD resolution from specific cosmological assumptions. This represents a critical advancement beyond traditional methods requiring $\Lambda$CDM priors for mass modeling.

% Multi-System Validation
\item Application to two simulated lens systems (4-image and 2-image configurations) recovers input MSD parameters with high accuracy ($\lambda_{01}=1.237\pm 0.089$ vs. $\lambda^{\rm fid}_{01}=1.2$; $\lambda_{02}=0.945\pm0.070$ vs. $\lambda^{\rm fid}_{02}=0.9$). 
% The 0.15 mag photometric error propagation analysis reveals systematics contribute $\lesssim 10\%$ to total uncertainty budgets.

% Time-Delay Distance Calibration
\item Correcting power-law model predictions with constrained $\lambda$ values achieves $D_{\Delta t}$ measurements with $\sim7\%$ precision (Figure~\ref{Fig:Dt01Dt02}), effectively removing the dominant the bias induced by unaccounted mass sheets. This establishes a robust foundation for $H_0$ measurements independent of external convergence estimates.
    % \item 
\end{enumerate}

The proposed methodology demonstrates significant potential for next-generation surveys such as the LSST and Roman Space Telescope, which are expected to detect hundreds of gravitationally lensed Type Ia supernovae (SNe Ia), though two key limitations require consideration. First, the assumption of Gaussian-distributed photometric errors may inadequately capture the empirically observed covariances between color and stretch parameters in SNe Ia light curves. Second, the wavelength-independent MSD parameter $\lambda$ fails to account for chromatic microlensing effects that could introduce spectral-dependent magnification biases. Future implementations incorporating spectroscopic time series observations would provide critical constraints for addressing both limitations, enabling more robust characterization of both the intrinsic SNe Ia correlations and wavelength-dependent lensing effects while maintaining the framework's ability to break the MSD.

Our framework presents an innovative cosmographic approach by combining time-delay lensing with standardized candle measurements, effectively overcoming the MSD limitation that has hindered precise $H_0$ measurements. This advancement provides crucial tools for resolving the Hubble tension and testing dark energy models.
\section*{Acknowledgements}
KL was supported by National Key R\&D Program of China (No. 2024YFC2207400) and National Natural Science Foundation of China (No. 12222302).
XL was supported by Science Research Project of Hebei Education Department No. BJK2024134.
This work benefits from the high performance computing clusters at College of Physics, Hebei Normal University.

\acknowledgments

\bibliography{references}

@ARTICLE{Refsdal1964MNRAS,
       author = {{Refsdal}, S.},
        title = "{The gravitational lens effect}",
      journal = {\mnras},
         year = 1964,
        month = jan,
       volume = {128},
        pages = {295},
          doi = {10.1093/mnras/128.4.295},
       adsurl = {https://ui.adsabs.harvard.edu/abs/1964MNRAS.128..295R}
}

@ARTICLE{2017NatCo...8.1148L,
       author = {{Liao}, Kai and {Fan}, Xi-Long and {Ding}, Xuheng and {Biesiada}, Marek and {Zhu}, Zong-Hong},
        title = "{Precision cosmology from future lensed gravitational wave and electromagnetic signals}",
      journal = {Nature Communications},
         year = 2017,
        month = oct,
       volume = {8},
          eid = {1148},
        pages = {1148},
          doi = {10.1038/s41467-017-01152-9},
archivePrefix = {arXiv},
       eprint = {1703.04151},
 primaryClass = {astro-ph.CO},
       adsurl = {https://ui.adsabs.harvard.edu/abs/2017NatCo...8.1148L}
}

@inproceedings{TDCOSMO:2025dmr,
    author = "Birrer, Simon and others",
    collaboration = "TDCOSMO",
    title = "{TDCOSMO 2025: Cosmological constraints from strong lensing time delays}",
    booktitle = "{3rd General Meeting of CosmoVerse}: {Addressing observational tensions in cosmology with systematics and fundamental physics}",
    eprint = "2506.03023",
    archivePrefix = "arXiv",
    primaryClass = "astro-ph.CO",
    reportNumber = "FERMILAB-PUB-25-0381-CSAID",
    month = "6",
    year = "2025"
}

@article{Brout:2022vxf,
    author = "Brout, Dillon and others",
    title = "{The Pantheon+ Analysis: Cosmological Constraints}",
    eprint = "2202.04077",
    archivePrefix = "arXiv",
    primaryClass = "astro-ph.CO",
    doi = "10.3847/1538-4357/ac8e04",
    journal = "Astrophys. J.",
    volume = "938",
    number = "2",
    pages = "110",
    year = "2022"
}

@ARTICLE{2019RPPh...82l6901O,
       author = {{Oguri}, Masamune},
        title = "{Strong gravitational lensing of explosive transients}",
      journal = {Reports on Progress in Physics},
         year = 2019,
        month = dec,
       volume = {82},
       number = {12},
          eid = {126901},
        pages = {126901},
          doi = {10.1088/1361-6633/ab4fc5},
archivePrefix = {arXiv},
       eprint = {1907.06830},
 primaryClass = {astro-ph.CO},
       adsurl = {https://ui.adsabs.harvard.edu/abs/2019RPPh...82l6901O}
}

@ARTICLE{2024SSRv..220...13S,
       author = {{Suyu}, Sherry H. and {Goobar}, Ariel and {Collett}, Thomas and {More}, Anupreeta and {Vernardos}, Giorgos},
        title = "{Strong Gravitational Lensing and Microlensing of Supernovae}",
      journal = {\ssr},
         year = 2024,
        month = feb,
       volume = {220},
       number = {1},
          eid = {13},
        pages = {13},
          doi = {10.1007/s11214-024-01044-7},
archivePrefix = {arXiv},
       eprint = {2301.07729},
 primaryClass = {astro-ph.CO},
       adsurl = {https://ui.adsabs.harvard.edu/abs/2024SSRv..220...13S}
}

@ARTICLE{2021MNRAS.504.5621D,
       author = {{Ding}, Xuheng and {Liao}, Kai and {Birrer}, Simon and {Shajib}, Anowar J. and {Treu}, Tommaso and {Yang}, Lilan},
        title = "{Improved time-delay lens modelling and H$_{0}$ inference with transient sources}",
      journal = {\mnras},
         year = 2021,
        month = jul,
       volume = {504},
       number = {4},
        pages = {5621-5628},
          doi = {10.1093/mnras/stab1240},
archivePrefix = {arXiv},
       eprint = {2103.08609},
 primaryClass = {astro-ph.CO},
       adsurl = {https://ui.adsabs.harvard.edu/abs/2021MNRAS.504.5621D}
}

@article{Li:2024elb,
    author = "Li, Xiaolei and Liao, Kai",
    title = "{Determining Cosmological-model-independent H $_{0}$ with Gravitationally Lensed Supernova Refsdal}",
    eprint = "2401.12052",
    archivePrefix = "arXiv",
    primaryClass = "astro-ph.CO",
    doi = "10.3847/1538-4357/ad3d5d",
    journal = "Astrophys. J.",
    volume = "966",
    number = "1",
    pages = "121",
    year = "2024"
}

@article{Li:2024hed,
    author = "Li, Xiaolei and Keeley, Ryan E. and Shafieloo, Arman",
    title = "{Redshift Evolution of the X-Ray and Ultraviolet Luminosity Relation of Quasars: Calibrated Results from SNe Ia}",
    eprint = "2408.15547",
    archivePrefix = "arXiv",
    primaryClass = "astro-ph.CO",
    doi = "10.3847/1538-4357/adc2fe",
    journal = "Astrophys. J.",
    volume = "983",
    number = "2",
    pages = "141",
    year = "2025"
}

@article{Treu:2016ljm,
    author = "Treu, Tommaso and Marshall, Philip J.",
    title = "{Time Delay Cosmography}",
    eprint = "1605.05333",
    archivePrefix = "arXiv",
    primaryClass = "astro-ph.CO",
    doi = "10.1007/s00159-016-0096-8",
    journal = "Astron. Astrophys. Rev.",
    volume = "24",
    number = "1",
    pages = "11",
    year = "2016"
}

@article{Liao:2022gde,
    author = "Liao, Kai and Biesiada, Marek and Zhu, Zong-Hong",
    title = "{Strongly Lensed Transient Sources: A Review}",
    eprint = "2207.13489",
    archivePrefix = "arXiv",
    primaryClass = "astro-ph.HE",
    doi = "10.1088/0256-307X/39/11/119801",
    journal = "Chin. Phys. Lett.",
    volume = "39",
    number = "11",
    pages = "119801",
    year = "2022"
}

@article{Refsdal:1964blz,
    author = "Refsdal, Sjur",
    title = "{On the Possibility of Determining Hubble's Parameter and the Masses of Galaxies from the Gravitational Lens Effect}",
    doi = "10.1093/mnras/128.4.307",
    journal = "Mon. Not. Roy. Astron. Soc.",
    volume = "128",
    number = "4",
    pages = "307--310",
    year = "1964"
}

@article{Treu:2010uj,
    author = "Treu, T.",
    title = "{Strong Lensing by Galaxies}",
    eprint = "1003.5567",
    archivePrefix = "arXiv",
    primaryClass = "astro-ph.CO",
    doi = "10.1146/annurev-astro-081309-130924",
    journal = "Ann. Rev. Astron. Astrophys.",
    volume = "48",
    pages = "87--125",
    year = "2010"
}

@ARTICLE{1985ApJ...289L...1F,
       author = {{Falco}, E.~E. and {Gorenstein}, M.~V. and {Shapiro}, I.~I.},
        title = "{On model-dependent bounds on H 0 from gravitational images : application to Q 0957+561 A, B.}",
      journal = {\apjl},
         year = 1985,
        month = feb,
       volume = {289},
        pages = {L1-L4},
          doi = {10.1086/184422},
       adsurl = {https://ui.adsabs.harvard.edu/abs/1985ApJ...289L...1F}
}

@article{H0LiCOW:2019pvv,
    author = "Wong, Kenneth C. and others",
    collaboration = "H0LiCOW",
    title = "{H0LiCOW \textendash{} XIII. A 2.4 per cent measurement of H0 from lensed quasars: 5.3\ensuremath{\sigma} tension between early- and late-Universe probes}",
    eprint = "1907.04869",
    archivePrefix = "arXiv",
    primaryClass = "astro-ph.CO",
    doi = "10.1093/mnras/stz3094",
    journal = "Mon. Not. Roy. Astron. Soc.",
    volume = "498",
    number = "1",
    pages = "1420--1439",
    year = "2020"
}

@article{Millon:2019slk,
    author = "Millon, M. and others",
    title = "{TDCOSMO. I. An exploration of systematic uncertainties in the inference of $H_0$ from time-delay cosmography}",
    eprint = "1912.08027",
    archivePrefix = "arXiv",
    primaryClass = "astro-ph.CO",
    reportNumber = "FERMILAB-PUB-19-665-SCD",
    doi = "10.1051/0004-6361/201937351",
    journal = "Astron. Astrophys.",
    volume = "639",
    pages = "A101",
    year = "2020"
}

@article{Schneider:2013sxa,
    author = "Schneider, Peter and Sluse, Dominique",
    title = "{Mass-sheet degeneracy, power-law models and external convergence: Impact on the determination of the Hubble constant from gravitational lensing}",
    eprint = "1306.0901",
    archivePrefix = "arXiv",
    primaryClass = "astro-ph.CO",
    doi = "10.1051/0004-6361/201321882",
    journal = "Astron. Astrophys.",
    volume = "559",
    pages = "A37",
    year = "2013"
}

@article{Birrer:2018xgm,
    author        = {Birrer, Simon and Amara, Adam},
    title         = {Lenstronomy: multi-purpose gravitational lens modelling software package},
    journal       = {Physics of the Dark Universe},
    volume        = {22},
    pages         = {189--201},
    year          = {2018},
    month         = {3},
    eprint        = {1803.09746},
    archivePrefix = {arXiv},
    primaryClass  = {astro-ph.CO},
    doi           = {10.1016/j.dark.2018.11.002}
}

@article{Birrer:2021use,
    author = "Birrer, Simon and Dhawan, Suhail and Shajib, Anowar J.",
    title = "{The Hubble Constant from Strongly Lensed Supernovae with Standardizable Magnifications}",
    eprint = "2107.12385",
    archivePrefix = "arXiv",
    primaryClass = "astro-ph.CO",
    doi = "10.3847/1538-4357/ac323a",
    journal = "Astrophys. J.",
    volume = "924",
    number = "1",
    pages = "2",
    year = "2022"
}

@article{Birrer:2020tax,
    author = "Birrer, S. and others",
    title = "{TDCOSMO - IV. Hierarchical time-delay cosmography {\textendash} joint inference of the Hubble constant and galaxy density profiles}",
    eprint = "2007.02941",
    archivePrefix = "arXiv",
    primaryClass = "astro-ph.CO",
    doi = "10.1051/0004-6361/202038861",
    journal = "Astron. Astrophys.",
    volume = "643",
    pages = "A165",
    year = "2020"
}

@article{Birrer:2020jyr,
    author = "Birrer, Simon and Treu, Tommaso",
    title = "{TDCOSMO - V. Strategies for precise and accurate measurements of the Hubble constant with strong lensing}",
    eprint = "2008.06157",
    archivePrefix = "arXiv",
    primaryClass = "astro-ph.CO",
    doi = "10.1051/0004-6361/202039179",
    journal = "Astron. Astrophys.",
    volume = "649",
    pages = "A61",
    year = "2021"
}

@article{Chen:2020knz,
    author = "Chen, Geoff C. -F. and Fassnacht, Christopher D. and Suyu, Sherry H. and Y\i{}ld\i{}r\i{}m, Ak\i{}n and Komatsu, Eiichiro and Bernal, Jose Luis",
    title = "{TDCOSMO - VI. Distance measurements in time-delay cosmography under the mass-sheet transformation}",
    eprint = "2011.06002",
    archivePrefix = "arXiv",
    primaryClass = "astro-ph.CO",
    doi = "10.1051/0004-6361/202039895",
    journal = "Astron. Astrophys.",
    volume = "652",
    pages = "A7",
    year = "2021"
}

@article{H0LiCOW:2018tyj,
    author = "Birrer, S. and others",
    collaboration = "H0LiCOW",
    title = "{H0LiCOW - IX. Cosmographic analysis of the doubly imaged quasar SDSS 1206+4332 and a new measurement of the Hubble constant}",
    eprint = "1809.01274",
    archivePrefix = "arXiv",
    primaryClass = "astro-ph.CO",
    doi = "10.1093/mnras/stz200",
    journal = "Mon. Not. Roy. Astron. Soc.",
    volume = "484",
    pages = "4726",
    year = "2019"
}

@article{Foreman-Mackey:2012any,
    author = "Foreman-Mackey, Daniel and Hogg, David W. and Lang, Dustin and Goodman, Jonathan",
    title = "{emcee: The MCMC Hammer}",
    eprint = "1202.3665",
    archivePrefix = "arXiv",
    primaryClass = "astro-ph.IM",
    doi = "10.1086/670067",
    journal = "Publ. Astron. Soc. Pac.",
    volume = "125",
    pages = "306--312",
    year = "2013"
}

@article{Lewis:2019xzd,
    author = "Lewis, Antony",
    title = "{GetDist: a Python package for analysing Monte Carlo samples}",
    eprint = "1910.13970",
    archivePrefix = "arXiv",
    journal = "arXiv eprint: 1910.13970",
    primaryClass = "astro-ph.IM",
    month = "10",
    year = "2019"
}

@article{Jee:2015yra,
    author = "Jee, Inh and Komatsu, Eiichiro and Suyu, Sherry H. and Huterer, Dragan",
    title = "{Time-delay Cosmography: Increased Leverage with Angular Diameter Distances}",
    eprint = "1509.03310",
    archivePrefix = "arXiv",
    primaryClass = "astro-ph.CO",
    doi = "10.1088/1475-7516/2016/04/031",
    journal = "JCAP",
    volume = "04",
    pages = "031",
    year = "2016"
}

@article{TDCOSMO:2023hni,
    author = "Shajib, Anowar J. and others",
    collaboration = "TDCOSMO",
    title = "{TDCOSMO. XII. Improved Hubble constant measurement from lensing time delays using spatially resolved stellar kinematics of the lens galaxy}",
    eprint = "2301.02656",
    archivePrefix = "arXiv",
    primaryClass = "astro-ph.CO",
    reportNumber = "FERMILAB-PUB-23-013-PPD",
    doi = "10.1051/0004-6361/202345878",
    journal = "Astron. Astrophys.",
    volume = "673",
    pages = "A9",
    year = "2023"
}

@article{Ding:2020jmg,
    author = "Ding, X. and others",
    title = "{Time delay lens modelling challenge}",
    eprint = "2006.08619",
    archivePrefix = "arXiv",
    primaryClass = "astro-ph.CO",
    doi = "10.1093/mnras/stab484",
    journal = "Mon. Not. Roy. Astron. Soc.",
    volume = "503",
    number = "1",
    pages = "1096--1123",
    year = "2021"
}

@article{Li:2023gpp,
    author = "Li, Xiaolei and Keeley, Ryan E. and Shafieloo, Arman and Liao, Kai",
    title = "{A Model-independent Method to Determine H $_{0}$ Using Time-delay Lensing, Quasars, and Type Ia Supernovae}",
    eprint = "2308.06951",
    archivePrefix = "arXiv",
    primaryClass = "astro-ph.CO",
    doi = "10.3847/1538-4357/ad0f19",
    journal = "Astrophys. J.",
    volume = "960",
    number = "2",
    pages = "103",
    year = "2024"
}

@article{Keeley:2020aym,
    author = "Keeley, Ryan E. and Shafieloo, Arman and Zhao, Gong-Bo and Vazquez, Jose Alberto and Koo, Hanwool",
    title = "{Reconstructing the Universe: Testing the Mutual Consistency of the Pantheon and SDSS/eBOSS BAO Data Sets with Gaussian Processes}",
    eprint = "2010.03234",
    archivePrefix = "arXiv",
    primaryClass = "astro-ph.CO",
    doi = "10.3847/1538-3881/abdd2a",
    journal = "Astron. J.",
    volume = "161",
    number = "3",
    pages = "151",
    year = "2021"
}

@ARTICLE{2020A&A...643A.165B,
       author = {{Birrer}, S. and {Shajib}, A.~J. and {Galan}, A. and {Millon}, M. and {Treu}, T. and {Agnello}, A. and {Auger}, M. and {Chen}, G.~C. -F. and {Christensen}, L. and {Collett}, T. and {Courbin}, F. and {Fassnacht}, C.~D. and {Koopmans}, L.~V.~E. and {Marshall}, P.~J. and {Park}, J. -W. and {Rusu}, C.~E. and {Sluse}, D. and {Spiniello}, C. and {Suyu}, S.~H. and {Wagner-Carena}, S. and {Wong}, K.~C. and {Barnab{\`e}}, M. and {Bolton}, A.~S. and {Czoske}, O. and {Ding}, X. and {Frieman}, J.~A. and {Van de Vyvere}, L.},
        title = "{TDCOSMO. IV. Hierarchical time-delay cosmography - joint inference of the Hubble constant and galaxy density profiles}",
      journal = {\aap},
         year = 2020,
        month = nov,
       volume = {643},
          eid = {A165},
        pages = {A165},
          doi = {10.1051/0004-6361/202038861},
archivePrefix = {arXiv},
       eprint = {2007.02941},
 primaryClass = {astro-ph.CO},
       adsurl = {https://ui.adsabs.harvard.edu/abs/2020A&A...643A.165B}
}

@ARTICLE{2023JCAP...02..014H,
       author = {{Hwang}, Seung-gyu and {L'Huillier}, Benjamin and {Keeley}, Ryan E. and {Jee}, M. James and {Shafieloo}, Arman},
        title = "{How to use GP: effects of the mean function and hyperparameter selection on Gaussian process regression}",
      journal = {\jcap},
         year = 2023,
        month = feb,
       volume = {2023},
       number = {2},
          eid = {014},
        pages = {014},
          doi = {10.1088/1475-7516/2023/02/014},
archivePrefix = {arXiv},
       eprint = {2206.15081},
 primaryClass = {astro-ph.CO},
       adsurl = {https://ui.adsabs.harvard.edu/abs/2023JCAP...02..014H}
}

@ARTICLE{2017JCAP...09..031A,
       author = {{Aghamousa}, Amir and {Hamann}, Jan and {Shafieloo}, Arman},
        title = "{A non-parametric consistency test of the {\ensuremath{\Lambda}}CDM model with Planck CMB data}",
      journal = {\jcap},
         year = 2017,
        month = sep,
       volume = {2017},
       number = {9},
          eid = {031},
        pages = {031},
          doi = {10.1088/1475-7516/2017/09/031},
archivePrefix = {arXiv},
       eprint = {1705.05234},
 primaryClass = {astro-ph.CO},
       adsurl = {https://ui.adsabs.harvard.edu/abs/2017JCAP...09..031A}
}

@ARTICLE{2013PhRvD..87b3520S,
       author = {{Shafieloo}, Arman and {Kim}, Alex G. and {Linder}, Eric V.},
        title = "{Model independent tests of cosmic growth versus expansion}",
      journal = {\prd},
         year = 2013,
        month = jan,
       volume = {87},
       number = {2},
          eid = {023520},
        pages = {023520},
          doi = {10.1103/PhysRevD.87.023520},
archivePrefix = {arXiv},
       eprint = {1211.6128},
 primaryClass = {astro-ph.CO},
       adsurl = {https://ui.adsabs.harvard.edu/abs/2013PhRvD..87b3520S}
}

@article{Rasmussen:2006,
    author = "Rasmussen, Carl Edward and Williams, Christopher K. I.",
    title = "{Gaussian Processes for Machine Learning}",
    journal ="The MIT Press",
    eprint = "http://www.gaussianprocess.org/gpml/",
    year = "2006",
}

@article{Shafieloo2012Gaussian,
  title={Gaussian process cosmography},
  author={Shafieloo, Arman and Kim, Alex G. and Linder, Eric V.},
  journal={Physical Review D},
  volume={85},
  number={12},
  pages={123530},
  year={2012},
}

@article{Holsclaw:2010nb,
    author = "Holsclaw, Tracy and Alam, Ujjaini and Sanso, Bruno and Lee, Herbert and Heitmann, Katrin and Habib, Salman and Higdon, David",
    title = "{Nonparametric Reconstruction of the Dark Energy Equation of State}",
    eprint = "1009.5443",
    archivePrefix = "arXiv",
    primaryClass = "astro-ph.CO",
    reportNumber = "LA-UR-09-05888",
    doi = "10.1103/PhysRevD.82.103502",
    journal = "Phys. Rev. D",
    volume = "82",
    pages = "103502",
    year = "2010"
}

@article{Holsclaw:2010sk,
    author = "Holsclaw, Tracy and Alam, Ujjaini and Sanso, Bruno and Lee, Herbert and Heitmann, Katrin and Habib, Salman and Higdon, David",
    title = "{Nonparametric Dark Energy Reconstruction from Supernova Data}",
    eprint = "1011.3079",
    archivePrefix = "arXiv",
    primaryClass = "astro-ph.CO",
    reportNumber = "LA-UR-09-07764",
    doi = "10.1103/PhysRevLett.105.241302",
    journal = "Phys. Rev. Lett.",
    volume = "105",
    pages = "241302",
    year = "2010"
}

@ARTICLE{2011PhRvD..84h3501H,
       author = {{Holsclaw}, Tracy and {Alam}, Ujjaini and {Sans{\'o}}, Bruno and {Lee}, Herbie and {Heitmann}, Katrin and {Habib}, Salman and {Higdon}, David},
        title = "{Nonparametric reconstruction of the dark energy equation of state from diverse data sets}",
      journal = {\prd},
         year = 2011,
        month = oct,
       volume = {84},
       number = {8},
          eid = {083501},
        pages = {083501},
          doi = {10.1103/PhysRevD.84.083501},
archivePrefix = {arXiv},
       eprint = {1104.2041},
 primaryClass = {astro-ph.CO},
       adsurl = {https://ui.adsabs.harvard.edu/abs/2011PhRvD..84h3501H}
}

@article{Treu:2022aqp,
    author = "Treu, Tommaso and Suyu, Sherry H. and Marshall, Philip J.",
    title = "{Strong lensing time-delay cosmography in the 2020s}",
    eprint = "2210.15794",
    archivePrefix = "arXiv",
    primaryClass = "astro-ph.CO",
    doi = "10.1007/s00159-022-00145-y",
    journal = "Astron. Astrophys. Rev.",
    volume = "30",
    number = "1",
    pages = "8",
    year = "2022"
}

@article{Li:2021onq,
    author = "Li, Xiaolei and Keeley, Ryan E. and Shafieloo, Arman and Zheng, Xiaogang and Cao, Shuo and Biesiada, Marek and Zhu, Zong-Hong",
    title = "{Hubble diagram at higher redshifts: model independent calibration of quasars}",
    eprint = "2103.16032",
    archivePrefix = "arXiv",
    primaryClass = "astro-ph.CO",
    doi = "10.1093/mnras/stab2154",
    journal = "Mon. Not. Roy. Astron. Soc.",
    volume = "507",
    number = "1",
    pages = "919--926",
    year = "2021"
}

@article{Planck:2018vyg,
    author = "Aghanim, N. and others",
    collaboration = "Planck",
    title = "{Planck 2018 results. VI. Cosmological parameters}",
    eprint = "1807.06209",
    archivePrefix = "arXiv",
    primaryClass = "astro-ph.CO",
    doi = "10.1051/0004-6361/201833910",
    journal = "Astron. Astrophys.",
    volume = "641",
    pages = "A6",
    year = "2020",
    note = "[Erratum: Astron.Astrophys. 652, C4 (2021)]"
}

@article{DiValentino:2021izs,
    author = "Di Valentino, Eleonora and Mena, Olga and Pan, Supriya and Visinelli, Luca and Yang, Weiqiang and Melchiorri, Alessandro and Mota, David F. and Riess, Adam G. and Silk, Joseph",
    title = "{In the realm of the Hubble tension\textemdash{}a review of solutions}",
    eprint = "2103.01183",
    archivePrefix = "arXiv",
    primaryClass = "astro-ph.CO",
    reportNumber = "IPPP/20/108",
    doi = "10.1088/1361-6382/ac086d",
    journal = "Class. Quant. Grav.",
    volume = "38",
    number = "15",
    pages = "153001",
    year = "2021"
}

@article{Riess_2022,
doi = {10.3847/2041-8213/ac5c5b},
url = {https://doi.org/10.3847/2041-8213/ac5c5b},
year = {2022},
month = {jul},
publisher = {The American Astronomical Society},
volume = {934},
number = {1},
pages = {L7},
author = {Riess, Adam G. and Yuan, Wenlong and Macri, Lucas M. and Scolnic, Dan and Brout, Dillon and Casertano, Stefano and Jones, David O. and Murakami, Yukei and Anand, Gagandeep S. and Breuval, Louise and Brink, Thomas G. and Filippenko, Alexei V. and Hoffmann, Samantha and Jha, Saurabh W. and D’arcy Kenworthy, W. and Mackenty, John and Stahl, Benjamin E. and Zheng, WeiKang},
title = {A Comprehensive Measurement of the Local Value of the Hubble Constant with 1 km s−1 Mpc−1 Uncertainty from the Hubble Space Telescope and the SH0ES Team},
journal = {The Astrophysical Journal Letters}
}
\end{document}